\documentstyle[aps,pra,epsfig,twocolumn]{revtex}

\def\be{\begin{equation}}
\def\ee{\end{equation}}
\def\bea{\begin{eqnarray}}
\def\eea{\end{eqnarray}}
\def\bma{\begin{mathletters}}
\def\ema{\end{mathletters}}
\def\tr{{\rm tr}}
\def\bk#1{\langle #1 \rangle}
\def\C{\hbox{$\mit I$\kern-.7em$\mit C$}}
\def\st{\mbox{ such that }}
\newcommand{\one}{\mbox{$1 \hspace{-1.0mm}  {\bf l}$}}
\begin{document}
\draft

\title{Optimization of entanglement witnesses}

\author{M. Lewenstein,$^{1}$ B. Kraus,$^{2}$ J. I. Cirac,$^{2}$
and P. Horodecki$^{3}$}

\address{$^1$ Institute for Theoretical Physics,
University of Hannover, D-30167 Hannover, Germany\\
 $^2$ Institute for
Theoretical Physics, University of Innsbruck, A--6020 Innsbruck,
Austria\\
 $^3$ Faculty of Applied Physics and Mathematics,
Technical University of Gda\'nsk, 80--952 Gda\'nsk, Poland}

\date{\today}

\maketitle
\begin{abstract}
An entanglement witness (EW) is an operator that allows to detect
entangled states. We give necessary and sufficient conditions for such
operators to be optimal, i.e. to detect entangled states in an optimal
way. We show how to optimize general EW, and then we particularize our
results to the non--decomposable ones; the latter are those that can
detect positive partial transpose entangled states (PPTES). We also
present a method to systematically construct and optimize this last
class of operators based on the existence of ``edge'' PPTES, i.e. states
that violate the range separability criterion [Phys. Lett. A{\bf 232},
333 (1997)] in an extreme manner. This method also permits the systematic construction of
non--decomposable positive maps (PM). Our results lead to a novel
sufficient condition for entanglement in terms of non-decomposable EW
and PM. Finally, we illustrate our results by constructing optimal EW
acting on $H=\C^2\otimes \C^4$. The corresponding PM constitute the first
examples of PM with minimal ``qubit'' domain, or -- equivalently --
minimal hermitian conjugate codomain.
\end{abstract}
\pacs{03.67.-a, 03.65.Bz, 03.65.Ca, 03.67.Hk}


\section{Introduction}

Quantum entanglement \cite{EPR,Sch}, which is an essence of many
fascinating quantum mechanical effects
\cite{Ekert,geste,Bennett_tel,comput}, is a very fragile phenomenon. It
is usually very hard to create, maintain, and manipulate entangled
states under laboratory conditions. In fact, any system is usually
subjected to the effects of external noise and interactions with the
environment. These effects turn {\it pure state entanglement} into {\it
mixed state}, or {\it noisy entanglement}. The separability problem,
that is, the characterization of mixed entangled states, is highly
nontrivial and has not been accomplished so far. Even the apparently
innocent question: {\it Is a given state entangled and does it contain
quantum correlations, or is it separable, and does not contain any
quantum correlations?} will, in general, be very hard (if not
impossible!) to answer.

Mathematically, mixed state entanglement can be described as follows. A
density operator $\rho\ge 0$ acting on a finite Hilbert space
$H=H_A\otimes H_B$ describing the state of two quantum systems {\it A} and {\it B}
is called separable \cite{We89} (or not entangled) if it can be written
as a convex combination of product vectors; that is,
in the form
\be
\label{rhosep}
\rho = \sum_k p_k |e_k,f_k\rangle\langle e_k,f_k|,
\ee
where $p_k\ge 0$, and $|e_k,f_k\rangle\equiv
|e_k\rangle_A\otimes|f_k\rangle_B$ are product vectors.
Conversely, $\rho$ is nonseparable (or entangled) if it cannot
be written in this form. Physically, a state described by a
separable (nonseparable) density operator $\rho$ can always
(never) be prepared locally. Most of the applications in quantum
information are based on the nonlocal properties of quantum
mechanics, \cite{Ekert,geste,Bennett_tel,comput,general} and
therefore on nonseparable states. Thus, a criterion to determine
whether a given density operator is nonseparable, i.e. useful
for quantum information purposes, or not is of crucial
importance. On the other hand, PPTES are objects of special
interest since they represent so--called bound entangled states,
and therefore provide an evidence of irreversibility in quantum
information processing \cite{Hohh}.

For low dimensional systems \cite{Pe96,Ho96} there exist operationally
simple necessary and sufficient conditions for separability. In fact, in
$H=\C^2\otimes\C^2$ and $H=\C^2\otimes\C^3$ the Peres--Horodecki
criterion \cite{Pe96,Ho96} establishes that $\rho$ is separable iff its
partial transpose is positive. Partial transpose means a transpose with
respect to one of the subsystems \cite{pt}. For higher dimensional
systems all operators with non--positive partial transposition are
entangled. However, there exist positive partial transpose entangled
states (PPTES) \cite{Ho97,Be98}.
Thus, the separability problem reduces to finding whether density operators with
positive partial transpose are separable or not \cite{primer,review}.

In the recent years there has been a growing effort in searching
for necessary and sufficient separability criteria and checks
which would be operationally simple \cite{primer,review}.
Several necessary \cite{We89,alfa} or sufficient
\cite{Ho97,karol,guifre,Kr99,Ho00} conditions for separability
are known. A particularly interesting necessary condition is
given by the so--called {\em range criterion} \cite{Ho97}.
According to this criterion, if the state $\rho$ acting on a
finite dimensional Hilbert space is separable then there must
exist a set of product vectors $\{|e_k,f_k\rangle\}$ that spans
the range $R(\rho)$ such that the set of partial complex
conjugated product states $\{|e_k,f^\ast_k\rangle\}$ spans the
range of the partial transpose of $\rho$ with respect to the
second system, i.e., $\rho^{T_B}$. Among the PPTES
that violate this criterion there are particular states
with the property that if one subtracts a projector
onto a product vector from them, the resulting operator is no
longer a PPTES \cite{Kr99,Ho00}. In this sense, these states lie
in the {\em edge} between PPTES and entangled states with
non--positive partial transposition, and therefore we will call
them ``edge'' PPTES. The analysis of the range of density
operators initiated in Ref. \cite{Ho97} has turned out to be very fruitful.  
In particular, it has led to an algorithm for the
optimal decomposition of mixed states into a separable and an
inseparable part \cite{Le98,Sa98,En00}, and to a systematic
method of constructing examples of PPTES using unextendible
product bases \cite{Be98,Te99}. For low rank operators it has
allowed to show that one can reduce the separability problem to
the one of determining the roots of certain complex polynomial
equations \cite{Kr99,Ho00} .

From a different point of view, a very general approach to
analyze the separability problem is based on the so--called
entanglement witnesses (EW) and positive maps (PM) \cite{Ho96}.
Entanglement witnesses \cite{Te99} are operators that detect the presence of
entanglement. Starting from these operators one can define PM's
\cite{Ja72} that also detect entanglement. An example of a PM is
precisely partial transposition \cite{Pe96,St63,Wo76}. The
importance of EW stems from the fact that a given operator is
separable iff there exists an EW that detects it \cite{Ho96}.
Thus, if one was able to construct all possible EW (or PM) one
would have solve the problem of separability. Unfortunately, it
is not known how to construct EW that detect PPTES in general.
The only result in this direction so far has been given in Ref.\
\cite{Te99}, although some preliminary results exist in the mathematical 
literature \cite{Ch82}. Starting from a PPTES fulfilling certain properties
(related to the existence of unextendible basis of product
vectors \cite{Be98}), it has been shown how to construct an EW
(and the corresponding PM) that detects it. Perhaps, one of the
most interesting goals regarding the separability problem is to
develop a constructive and operational approach using EW and PM
that allows us to detect mixed entanglement.

In this paper we realize this goal partially: we introduce a powerful
technique to construct EW and PM that, among other things, allows us to
study the separability of certain density operators . In particular, we
show how to construct optimal EW; that is, operators that detect the
presence of entanglement in an optimal way. We specifically concentrate
on non--decomposable EW, which are those that detect the presence of
PPTES. Furthermore, we present a way of constructing optimal EW for edge
PPTES. Our method generalizes the one introduced by Terhal \cite{Te99}
to the case in which there are no unextendible basis of product vectors.
When combined with our previous results \cite{Kr99,Ho00} regarding
subtracting product vectors from PPTES, the construction of
non--decomposable optimal EW starting from ``edge'' PPTES gives rise to a
novel sufficient criterion for non--separability of general density
operators with positive partial transposition. We illustrate our method
by constructing optimal EW that detect some known examples of PPTES
\cite{Ho97} in $H=\C^2\otimes \C^4$. The corresponding PM constitute the
first examples of PM with minimal ``qubit'' domain, or -- equivalently
-- minimal hermitian conjugate codomain.

This paper is organized as follows. In Section II we review the
definition of EW and fix some notation. In Section III we study
general EW. We define optimal witnesses and find a criterion to
decide whether an EW is optimal or not. In Section IV we
restrict the results of Section III to non--decomposable EW. In
particular, we show how to optimize them by subtracting
decomposable operators. In Section V we give an explicit method
to optimize both, general and non--decomposable EW. We also show
how to construct non--decomposable EW, and that this leads to a
sufficient criterion of non--separability. The construction and
optimization is based on the use of ``edge'' PPTES. In Section
VI we extend our results to positive maps. In Section VII we
illustrate our methods and results starting from the examples of
PPTES given in Ref. \cite{Ho97}. The paper also contains two
appendices. In Appendix A we describe in detail a method to
check whether an EW is optimal or not. In Appendix B we discuss
separately some important properties of the edge PPTES, and show
that they provide a canonical decomposition of mixed states with
positive partial transpose.

\section{Definitions and notation}

We say that an operator $W=W^\dagger$ acting on $H=H_A\otimes
H_B$ is an EW if  \cite{Ho96,Te99}:

\begin{description}

\item [(I)] $\langle e,f|W|e,f\rangle\ge 0$ for all product vectors
$|e,f\rangle$;

\item [(II)] has at least one negative eigenvalue (i.e. is not positive);

\item [(III)] $\tr(W)=1$.

\end{description}

The first property (I) implies that $\langle \rho\rangle_W \equiv
\tr(W\rho) \ge 0$ for all $\rho$ separable. Thus, if we have $\langle
\rho\rangle_W<0$ for some $\rho\ge 0$, then $\rho$ is nonseparable. In
that case we say that $W$ {\em detects} $\rho$. The second one (II)
implies that every EW detects something, since in particular it detects
the projector on the subspace corresponding to the negative eigenvalues
of $W$. The third property (III) is just normalization condition that
we need in order to compare the action of different EW \cite{note0}.

In this paper we will denote by $K(\rho)$ and $R(\rho)$ the
kernel and range of $\rho$, respectively. The partial
transposition of an operator $X$ will be denoted by $X^T$
\cite{pt,pt2}. On the other hand, we will encounter several
kinds of operators (EW, positive operators, decomposable
operators, etc) and vectors. In order to help to identify the
kind of operators and vectors we use, and not to overwhelm the
reader by specifying at each point their properties, we will use
the following notation:

\begin{itemize}

\item $W$ will denote an EW.

\item $P,Q$ will denote positive operators. Unless specified
they will have unit trace [$\tr(P)=\tr(Q)=1$].

\item $D$ will denote a decomposable operator. That is, $D=aP+bQ^T$,
where $a,b\ge 0$. Unless stated, all decomposable operators that we use
will have unit trace (i.e., $b=1-a$).

\item $\rho$ will denote a positive operator (not necessarily of trace 1).

\item $|e,f\rangle$ will denote product vectors with $|e\rangle\in H_A$
and $|f\rangle\in H_B$. Unless especified, they will be normalized.

\end{itemize}

\section{General entanglement witnesses}

In this Section we first give some definitions directly related to EW.
Then we introduce the concept of optimal EW. We derive a criterion to
determine when an EW is optimal. This criterion will serve us to find an
optimization procedure for these operators.

\subsection{Definitions}

Given an EW, $W$, we define:

\begin{itemize}

\item $D_W=\{\rho\ge 0, \st
\bk{\rho}_W<0\}$; that is, the set of operators detected by $W$.

\item Finer: Given two EW, $W_1$ and $W_2$, we say that $W_2$ is {\em finer}
than $W_1$, if $D_{W_1}\subseteq D_{W_2}$; that is, if all the
operators detected by $W_1$ are also detected by $W_2$.

\item Optimal entanglement witness (OEW): We say that $W$ is an OEW if
there exist no other EW which is finer.

\item $P_W=\{|e,f\rangle\in H, \st \langle
e,f|W|e,f\rangle=0\}$; that is, the set of product vectors on which $W$
vanishes. As we will show, these vectors are closely related to
the optimality property.

\end{itemize}

Note the important role that the vectors in $P_W$ play regarding
entanglement (for a method to determine $P_W$ in practice, see Appendix A).
If we have an EW, $W$, which detects a given
operator $\rho$, then the operator $\rho'=\rho+\rho_w$ where
\be
\label{rhow}
\rho_w = \sum_k p_k |e_k,f_k\rangle\langle e_k,f_k|
\ee
with $p_k\ge 0$, and $|e_k,f_k\rangle\in P_W$ is also detected
by $W$. In fact, this means that any operator of the form
(\ref{rhow}) is in the border between separable states and
non--separable states, in the sense that if we add an
arbitrarily small amount of $\rho$ to it we obtain a
non--separable state. Thus, the structure of the sets $P_W$
characterizes the border between separable and non--separable
states. In fact, from the results of this Section it will become
clear that we can restrict ourselves to the structure of the sets
of $P_W$ corresponding to OEW's.

\subsection{Optimal entanglement witnesses}

According to Ref.\ \cite{Ho96} $\rho$ is nonseparable iff there
exists an EW which detects it. Obviously, we can restrict
ourselves to the study of OEW. For that, we need criteria to
determine when an EW is optimal. In this subsection we will
derive a necessary and sufficient condition for this to happen
(Theorem 1 below). In order to do that, we first have to
introduce some results that tell us under which conditions an EW
is finer than another one.

{\bf Lemma 1:} Let $W_2$ be finer than $W_1$ and
\be
\label{lambda}
\lambda\equiv \inf_{\rho_1\in D_{W_1}} \left|
\frac{\bk{\rho_1}_{W_2}}{\bk{\rho_1}_{W_1}} \right|.
\ee
Then we have:

\begin{description}
\item [(i)] If $\bk{\rho}_{W_1}=0$ then $\bk{\rho}_{W_2}\le 0$.

\item  [(ii)] If $\bk{\rho}_{W_1}<0$, then $\bk{\rho}_{W_2} \le
\bk{\rho}_{W_1}$.

\item [(iii)] If $\bk{\rho}_{W_1}>0$ then $\lambda\bk{\rho}_{W_1}\ge
\bk{\rho}_{W_2}$.

\item [(iv)] $\lambda\ge 1$. In particular,
$\lambda=1$ iff $W_1=W_2$.

\end{description}

{\em Proof:} Since $W_2$ is finer than $W_1$ we will use the
fact that for all $\rho\ge 0$ such that $\bk{\rho}_{W_1}<0$ then
$\bk{\rho}_{W_2}<0$.

{\em (i)} Let us assume that $\bk{\rho}_{W_2} > 0$. Then we take
any $\rho_1\in {\cal D}_{W_1}$ so that for all $x\ge 0$, $0\le
\tilde \rho(x)\equiv \rho_1+x\rho \in {\cal D}_{W_1}$. But for
sufficiently large $x$ we have that
$\bk{\tilde \rho(x)}_{W_2}$
is positive, which cannot be since then $\rho(x)
\not{\in} {\cal D}_{W_2}$.

{\em (ii)} We define $\tilde \rho=\rho+|\bk{\rho}_{W_1}|\one\ge
0$. We have that $\bk{\tilde \rho}_{W_1}=0$. Using (i) we have
that $0\ge
\bk{\rho}_{W_2}+|\bk{\rho}_{W_1}|$.

{\em (iii)} We take $\rho_1\in D_{W_1}$ and define $\tilde \rho=
\bk{\rho}_{W_1} \rho_1 + |\bk{\rho_1}_{W_1}| \rho \ge 0$, so that $\bk{\tilde
\rho}_{W_1}=0$. Using (i) we have $|\bk{\rho_1}_{W_1}|\bk{\rho}_{W_2}\le
|\bk{\rho_1}_{W_2}|\bk{\rho}_{W_1}$. Dividing both sides by
$|\bk{\rho_1}_{W_1}|>0$ and $\bk{\rho}_{W_1}>0$ we obtain
\be
\frac{\bk{\rho}_{W_2}}{\bk{\rho}_{W_1}} \le  \left|
\frac{\bk{\rho_1}_{W_2}}{\bk{\rho_1}_{W_1}} \right|.
\ee
Taking the infimum with respect to $\rho_1\in D_{W_1}$ in the
rhs of this equation we obtain the desired result.

{\em (iv)} From (ii) immediately follows that $\lambda\ge 1$. On
the other hand, we just have to prove that if $\lambda=1$ then
$W_1=W_2$ (the only if part is trivial). If $\lambda=1$, using
(i) and (iii) we have that
$\bk{\rho_v}_{W_1}\ge\bk{\rho_v}_{W_2}$ for all
$\rho_v=|e,f\rangle\langle e,f|$ projector on a product vector.
Since $\tr(W_1)=\tr(W_2)$ we must have $\tr[(W_1-W_2)\rho_v]=0$
for all $\rho_v$, since we can always find a product basis in
which we can take the trace. But now, for any given $\rho\ge 0$ we
can define $\tilde
\rho(x)=\rho+x \one$ such that for large enough $x$, $\tilde \rho(x)$ is
separable \cite{karol}. In that case we have $\bk{\tilde
\rho(x)}_{W_1}=\bk{\tilde \rho(x)}_{W_2}$ which implies that
$\bk{\rho}_{W_1}=\bk{\rho}_{W_2}$, i.e. $W_1=W_2$. $\Box$

{\bf Corollary 1:} $D_{W_1}=D_{W_2}$ iff $W_1=W_2$.

{\it Proof:} We just have to prove the only if part. For that,
we define $\lambda$ as in (\ref{lambda}). On the other hand,
defining
\be
\tilde\lambda\equiv \inf_{\rho_2\in D_{W_2}} \left|
\frac{\bk{\rho_2}_{W_1}}{\bk{\rho_2}_{W_2}}\right|
\ee
we have that $\tilde\lambda\ge 1$ since $W_1$ is finer than
$W_2$ (Lemma 1(iv)). Equivalently,
\be
1\ge \sup_{\rho_1\in D_{W_1}} \left|
\frac{\bk{\rho_1}_{W_2}}{\bk{\rho_1}_{W_1}}\right| \ge \lambda\ge 1,
\ee
where for the last inequality we have used that $W_2$ is finer
than $W_1$. Now, since $\lambda=1$ we have that $W_1=W_2$
according to Lemma 1(iv). $\Box$

Next, we introduce one of the basic results of this paper. It basically
tell us that EW is finer than another one if they differ by a positive
operator. That is, if we have an EW and we want to find another one
which is finer, we have to subtract a positive operator.

{\bf Lemma 2:} $W_2$ is finer than $W_1$ iff there exists a $P$ and
$1>\epsilon\ge 0$ such that $W_1=(1-\epsilon)W_2+\epsilon P$.

{\it Proof:} (If) For all $\rho\in D_{W_1}$ we have that
$0>\bk{\rho}_{W_1}= (1-\epsilon)\bk{\rho}_{W_2}+\epsilon \bk{\rho}_{P}$
which implies $\bk{\rho}_{W_2}<0$ and therefore $\rho\in D_{W_2}$. (Only
if) We define $\lambda$ as in (\ref{lambda}). Using Lemma 1(iv) we have
$\lambda\ge 1$. First, if $\lambda=1$ then according to Lemma 1(iv) we
have $W_1=W_2$ (i.e., $\epsilon=0$). For $\lambda> 1$, we define
$P=(\lambda-1)^{-1}( \lambda W_1-W_2)$ and $\epsilon=1-1/\lambda>0$.
We have that $W_1=(1-\epsilon)W_2+\epsilon P$, so that it only remains
to be shown that $P\ge 0$. But this follows from Lemma 1(i--iii) and the 
definition of $\lambda$, $\lambda=\inf_{\rho_1\in D_{W_1}} \left|
\frac{\bk{\rho_1}_{W_2}}{\bk{\rho_1}_{W_1}} \right|$
. $\Box$

The previous lemma provides us with a way of determining when an
EW is finer than another one. With this result, we are now at
the position of fully characterizing OEW.

{\bf Theorem 1:} $W$ is optimal iff for all $P$ and
$\epsilon>0$, $W'=(1+\epsilon)W-\epsilon P$ is not an EW [does
not fulfill (I)].

{\em Proof:} (If) According to Lemma 2, there is no EW which is
finer than $W$, and therefore $W$ is optimal. (Only if) If $W'$
is an EW, then according to Lemma 2 $W$ is not optimal. $\Box$

The previous theorem tells us that $W$ is optimal iff when we
subtract any positive operator from it, the resulting operator
is not positive on product vectors. This result is not very
practical because of two reasons: (1) for a given $P$ it is
typically very hard to check whether there exists some
$\epsilon>0$ such that $W-\epsilon P$ is positive on all product
vectors; (2) it may be difficult to find a particular $P$ that
can be subtracted from $W$ among all possible positive
operators. In Appendix A we show how to circumvent these two
drawbacks in practice: we give a simple criterion to determine
when a given $P$ can be subtracted from $W$. This allows us to
determine which are the positive operators which can be
subtracted from a given EW.

In the rest of this subsection we will present some simple results
related to these two questions. First, it is clear that not every
positive operator $P$ can be subtracted from an EW, $W$. In particular,
the following lemma tells us that it must vanish on $P_W$.

{\bf Lemma 3:} If $PP_W\ne 0$ then $P$ cannot be subtracted
from $W$.

{\em Proof:} There exists some $|e_0,f_0\rangle\in P_W$ such
that $\langle e_0,f_0|P|e_0,f_0\rangle >0$. Substituting this
product vector in the condition I for any $W-\epsilon P$ we see
that the inequality is not fulfilled for any $\epsilon>0$, i.e.
$P$ cannot be subtracted. $\Box$

{\bf Corollary 2:} If $P_W$ spans $H$ then $W$ is optimal.

Note that, as announced at the beginning of this Section, the set $P_W$
plays an important role in determining the properties of the separable
states which lie on the border with the entangled states. We see here,
that this set also plays an important role in determining whether an EW
is optimal or not.

On the other hand, in order to check whether a given operator $P$ can be
subtracted or not from $W$, one has to check whether there exists some
$\epsilon >0$ such that $\langle e,f|W-\epsilon P|e,f\rangle>0$ for all
$|e,f\rangle$. The following lemma gives an alternative way to do this.
In fact, it gives a necessary and sufficient criterion for an EW to be
optimal. For a given $|e\rangle \in H_A$, we will denote by
$W_e \equiv\langle e|W|e\rangle$.

{\bf Lemma 4:} $W$ is optimal iff for all $|\Psi\rangle$
orthogonal to $P_W$
\be
\epsilon\equiv \inf_{|e\rangle\in H_A} \left[
\langle\Psi|e\rangle W_e^{-1}\langle e|\Psi\rangle
\right]^{-1} =0.
\ee

{\em Proof:} (If) Let us assume that $W$ is not optimal; that
is, there exists $W'\ne W$, finer than $W$. Then,
according to Lemma 2 we have that there exists $\epsilon_0>0$
and $P\ge 0$ such that $W'= (W-\epsilon_0P)/(1-\epsilon_0)$.
Imposing that $W'$ is positive on product vectors (i.e. $W_e'\ge 0$
for all $|e\rangle \in H_A$) we obtain $0\le
\langle e|W-\epsilon_0P|e\rangle\le W_e
-\epsilon_0 \lambda_{\Psi}\langle e|\Psi\rangle\langle \Psi|
e\rangle$, where $|\Psi\rangle$ is any eigenstate of $P$ with
nonzero eigenvalue $\lambda_{\Psi}$. According to Ref.\ \cite{Kr99}, this last
operator is positive iff both: (i) $\langle e|\Psi\rangle$ is in
the range of $\langle e|W|e\rangle$, which imposes that
$|\Psi\rangle$ is orthogonal to $P_W$; (ii) $\lambda_{\Psi}\epsilon_0\le
\left[\langle\Psi|e\rangle W_e^{-1}\langle e|\Psi\rangle \right]^{-1}$, which
imposes that $\epsilon\ge\lambda_{\Psi}\epsilon_0>0$ for that given
$|\Psi\rangle$. (Only if) Let us assume that there exists some
$|\Psi\rangle$ orthogonal to $P_W$ such that $\epsilon>0$. Then,
using the same arguments one can show that $W'\equiv
(W-\epsilon|\Psi\rangle\langle\Psi|)/(1-\epsilon)\ne W$ is an
EW. According to Lemma 2, $W'$ is finer than $W$, so that $W$ is
not optimal. $\Box$.

\subsection{Decomposable entanglement witnesses}

There exists a class of EW which is very simple to characterize,
namely the decomposable entanglement witnesses (d--EW)
\cite{Wo76}. Those are EW that can be written in the form
\be
\label{DEW}
W=a P+ (1-a) Q^T,
\ee
where $a\in [0,1]$. As it is well known (see next section),
these EW cannot detect PPTES. In any case, for the sake of
completeness, we will give some simple properties of optimal
d--EW.

{\bf Theorem 2:} Given a d--EW, $W$, if it is optimal then it
can be written as $W=Q^T$, where $Q\ge 0$ contains no product
vector in its range.

{\em Proof:} Since $W$ is decomposable, it can be written as
$W=aP+(1-a)Q^T$. $W'\propto W-aP$ is also a witness, which
according to Lemma 2 is finer than $W$, and therefore $W$ is not
optimal. On the other hand, if $|e,f\rangle\in R(Q)$ then for
some $\lambda>0$ we have that $W\propto
(Q-\lambda|e,f\rangle\langle e,f|)^T$ is finer than $W$, and
therefore this last is not optimal. $\Box$

This previous result can be slightly generalized as follows:

{\bf Theorem 2':} Given a d--EW, $W$, if it is optimal then it
can be written as $W=Q^T$, where $Q\ge 0$ and there is no
operator $P\in R(Q)$ such that $P^T\ge 0$.

{\em Proof:} Is the same as in previous theorem. $\Box$

{\bf Corollary 3:} Given a d--EW, $W$, if it is optimal then
$W^{T}$ is not an EW [does not fulfill (II)].

{\em Proof:} Using Theorem 2 we have that $W=Q^T$ with $Q\ge 0$.
Then $W^T=Q\ge 0$, which does not satisfy property (ii). $\Box$

\section{Non--decomposable entanglement witnesses}

In the previous section we have been concerned with EW in
general. As mentioned above, when studying separability we just
have to consider those EW that can detect PPTES. In
order to characterize them, one defines non--decomposable
witnesses (nd--EW) as those EW which cannot be written in the
form (\ref{DEW}) \cite{Wo76}. This Section is devoted to this
kind of witnesses. The importance of nd--EW in order to detect
PPTES is reflected in the following

{\bf Theorem 3:} An EW is non--decomposable iff it detects
PPTES.

{\em Proof:} (If) Let us assume that the EW is decomposable.
Then it cannot detect PPT, since if $\rho,\rho^T\ge 0$ we have
$\tr[(aP+(1-a)Q^T)\rho] =a\tr(P\rho)+(1-a)\tr(Q\rho^T)\ge 0$.
(Only if) The set of decomposable witnesses is convex and
closed, and $W$, as a set containing one point, is a closed convex
 set itself. Thus, from
Hahn--Banach theorem \cite{Ru73} it follows that there exists an
operator $\rho$ such that: (i) $\tr[\rho(aP+(1-a)Q^T)]\geq 0$ for
all $P,Q\ge 0$, $a\in[0,1]$; (ii) $\tr(\rho W)<0$. From (i),
taking $a=1$ we infer that $\rho\ge 0$; on the other hand,
taking $a=0$ we obtain that $\tr[\rho^TQ]\ge 0$ for all $Q\ge
0$, and therefore $\rho^T\ge 0$. Thus, $W$ detects $\rho$ which
is a PPTES. $\Box$

{\bf Corollary 4:} Given an operator $D$, it is decomposable iff
$\tr(D\rho)\ge 0$ for all $\rho,\rho^T\ge 0$.

\subsection{Definitions}

In this Subsection we introduce some definitions which are
parallel to those given in the previous Section. Given a nd--EW,
$W$, we define:

\begin{itemize}

\item $d_W=\{\rho\ge 0, \st \rho^T\ge 0 {\mbox{
and }} \bk{\rho}_W<0\}$; that is, the set of PPT operators
detected by $W$.

\item Non--decomposable-finer (nd--finer): Given two nd--EW, $W_1$ and
$W_2$, we say that $W_2$ is {\em nd--finer} than $W_1$, if
$d_{W_1}\subseteq d_{W_2}$; that is, if all the operators detected by
$W_1$ are also detected by $W_2$.

\item Non--decomposable optimal entanglement witness (nd--OEW): We say
that $W$ is an nd--OEW if there exist no other nd--EW which is
nd--finer.

\item $p_W=\{|e,f\rangle\in H, \st \langle e,f|W|e,f\rangle=0\}$; that
is, the product vectors on which $W$ vanishes.

\end{itemize}

Note again the important role that the vectors in $p_W$ play regarding
PPTES. If we have a nd--EW, $W$, which detects a given PPTES $\rho$,
then the operator $\rho'=\rho+\rho_w$ where $\rho_w$ has the form
(\ref{rhow}) with $p_k\ge 0$, and $|e_k,f_k\rangle\in p_W$ also
describes a PPTES. Thus, any operator of the form (\ref{rhow}) lies in
the border between separable states and PPTES.

\subsection{Optimal non--decomposable entanglement witness}

The goal of this section is to find a necessary and sufficient
condition for a nd--EW to be optimal. We start by proving a
similar result to the one given in Lemma 1, but for nd--EW:

{\bf Lemma 1b:} Let $W_2$ be nd--finer than $W_1$,
\be
\label{lambdand}
\lambda\equiv \inf_{\rho_1\in d_{W_1}} \left|
\frac{\tr(W_2 \rho_1)}{\tr(W_1 \rho_1)} \right|,
\ee
and now both, $\rho,\rho^T\ge 0$. Then we have have (i--iv) as in Lemma 1.

{\em Proof:} The proof is basically the same as in Lemma 1 and
will be omitted here.

{\bf Corollary 1b:} Given two nd--EW, $W_{1,2}$, then
$d_{W_1}=d_{W_2}$ iff $W_1=W_2$.

{\em Proof:} The proof is basically the same as Corollary 1 and
will be omitted here.

{\bf Lemma 2b:} Given two nd--EW, $W_{1,2}$, $W_2$ is nd--finer
than $W_1$ iff there exists a decomposable operator $D$ and
$1>\epsilon\ge 0$ such that $W_1=(1-\epsilon) W_2+\epsilon D$.

{\it Proof:} (If) Given any $\rho,\rho^T\ge 0$, we have that if
$\rho\in d_{W_1}$ then
$0>\bk{\rho}_{W_1}=(1-\epsilon)\bk{\rho}_{W_2} + \epsilon
\bk{\rho}_{D}\ge \bk{\rho}_{W_2}$, where in the last inequality we have
used that $\bk{\rho}_D\ge 0$ since $D$ is decomposable (see
Corollary 4). Therefore $\rho\in d_{W_2}$. (Only if) We define
$\lambda$ as in (\ref{lambdand}), so that $\lambda\ge 1$
according to Lemma 1b(iv). If $\lambda=1$ we have $W_1=W_2$. If
$\lambda>1$ we define $D=(\lambda-1)^{-1}(\lambda W_1-W_2)$ and
$\epsilon=1-1/\lambda$. We have that
$W_1=(1-\epsilon)W_2+\epsilon D$, so that it only remains to be
shown that $D$ is decomposable. But from Lemma 1b(i--iii) and the 
definition of $\lambda$ it
follows that $\bk{\rho}_D\ge 0$ for all $\rho,\rho^T\ge 0$.
Using Corollary 4 we then have that $D$ is decomposable. $\Box$.

Now we are able to fully characterize nd--OEW.

{\bf Theorem 1b:} Given an nd--EW, $W$, it is nd--optimal iff
for all decomposable operators $D$ and $\epsilon>0$,
$W'=(1+\epsilon) W-\epsilon D$ is not an EW [does not fulfill
(I)].

{\em Proof:} Is the same as for Theorem 1. $\Box$

Theorems 1 and 1b allow us to relate OEW and nd--OEW. In this
way we can directly translate the results for general OEW to
nd--OEW. We have

{\bf Theorem 4:} Given a nd--EW, $W$, $W$ is a nd--OEW iff both
$W$ and $W^T$ are OEW.

{\em Proof:} (If) Let us assume that $W$ is not a nd--OEW. Then,
according to Theorem 1b there exists $\epsilon>0$ and a
decomposable operator $D$ such that $W'=(1+\epsilon)W-\epsilon
D$ is a nd--EW. We can write $D=aP+(1-a)Q^T$, with $a\in[0,1]$.
If $a\ne 0$, then $W_1=(1+a\epsilon)W-a\epsilon P$ fulfills
$\langle e,f|W_1|e,f\rangle\ge
\langle e,f|W'|e,f\rangle\ge 0$, and therefore, according to Lemma 2,
$W$ is not optimal. If $a\ne 1$ then
$W_2=[1+(1-a)\epsilon]W^T-(1-a)\epsilon Q$ fulfills $\langle
e,f|W_2|e,f\rangle\ge \langle e,f|(W')^T|e,f\rangle\ge 0$, i.e.
is an EW and therefore $W^T$ is not optimal. (Only if) According
to Theorem 1b, if $W$ is nd--optimal then for all
$D=aP+(1-a)Q^T$, with $a\in[0,1]$, and all $\epsilon>0$ we have
that $W'=(1-\epsilon)W-\epsilon D$ does not satisfy (I).
Taking $a=1$ we have for all $P$ and $\epsilon>0$,
$W_1=(1-\epsilon)W-\epsilon P$ does not fulfill (I), and
therefore (Theorem 1) $W$ is optimal; analogously, taking $a=0$
we have that $W^T$ is optimal also. $\Box$

{\bf Corollary 5:} $W$ is a nd--OEW iff $W^T$ is an nd--OEW.

\section{Optimization}

In this Section we give a procedure to optimize EW which is
based on the results of the previous Sections.

\subsection{Optimization of general entanglement witnesses}

Our method is based in the following lemma. It tells us how much
we can subtract from an EW. Here we will denote by $W_e=\langle
e|W|e\rangle$ and $P_e=\langle e|P|e\rangle$ where $|e\rangle\in
H_A$, by $[\ldots]_{\rm min}$ the minimum eigenvalue, and by
$[\ldots]_{\rm max}$ the maximum eigenvalue. On the other hand,
$X^{-1/2}$ will denote the square root of the pseudoinverse of
$X$ \cite{notepseudoinv}.

{\bf Lemma 5:} If there exists some $P$ such that $PP_W=0$ and
\bea
\label{lambdap}
\lambda_0&\equiv&
\inf_{|e\rangle\in H_A} \left[P_e^{-1/2} W_e P_e^{-1/2}\right]_{\rm min} \\ \nonumber
&=&\left(\sup_{|e\rangle\in H_A} \left[W_e^{-1/2} P_e
W_e^{-1/2}\right]_{\rm max}\right)^{-1}>0.
\eea
then
\bea
W'(\lambda)\equiv (W-\lambda P)/(1-\lambda) 
\eea
with $\lambda>0$ is an EW iff $\lambda\le \lambda_0$.

{\em Proof:} Let us find out for which values of $\lambda\ge 0$,
$W'(\lambda)$ is an EW. We have to impose condition (I), which
can be written as $\langle e|W'(\lambda)|e\rangle\ge 0$, i.e.
\be
\label{We}
W_e - \lambda P_e \ge 0.
\ee
Multiplying by $P_e^{-1/2}$ on the right and left of this
equation we obtain $P_e^{-1/2} W_e P_e^{-1/2}\ge \lambda$, which
immediately gives that $\lambda\le \lambda_0$ given in the first
part of Eq.\ (\ref{lambdap}). On the other hand, multiplying by
$W_e^{-1/2}$ on the right and left of Eq.\ (\ref{We}) we obtain
$W_e^{-1/2} P_e W_e^{-1/2}\le 1/\lambda$, which immediately
gives that $\lambda\le
\lambda_0$ given in the second equality of Eq.\ (\ref{lambdap}).
$\Box$

Lemma 5 provides us with a direct method to optimize EW by
subtracting positive operators for which the elements of $P_W$
are contained in their kernels. The method thus consists of: (1)
determining $P_W$; (2) choosing an operator $P$ so that $PP_W=0$
and determining $\lambda$ using (\ref{lambdap}); (3) if
$\lambda\ne 0$ then we subtract the operator $P$ according to
Lemma 5. Continuing in the same vein we will reach an OEW. In
Appendix A we show how to accomplish steps (1) and (2) in
practice.

\subsection{Optimization of non--decomposable entanglement witnesses}

For nd--EW we have the following generalization of Lemma 5:

{\bf Lemma 5b:} Given a nd--EW, $W$, if there exists some
decomposable operator $D$ such that $Dp_W=0$ and
\bea
\label{epsilon0}
\lambda_0\equiv
\inf_{|e\rangle\in H_A} \left[D_e^{-1/2} W_e D_e^{-1/2}\right]_{\rm min} =\\ \nonumber
\left(\sup_{|e\rangle\in H_A}
\left[W_e^{-1/2} D_e W_e^{-1/2}\right]_{\rm max}\right)^{-1}>0.
\eea
then 
\bea
W'(\lambda)\equiv (W-\lambda D)/(1-\lambda)
\eea
 with $\lambda>0$ is a nd--EW iff $\lambda\le \lambda_0$.

{\em Proof:} Is the same as for Lemma 5.

With the help of Lemma 5b we can optimize nd--EW by subtracting
decomposable operators as follows: (1) determining $p_W$ and
$p_{W^T}$; (2) choosing $P,Q$ so that $Pp_W=0$ and $Qp_{W^T}=0$,
building $D=aP+(1-a)Q^T$ with $a\in[0,1]$, and determining
$\lambda_0$ using (\ref{epsilon0}); (3) if $\lambda_0\ne 0$ then
we subtract the operator $D$ according to Lemma 5b.

\subsection{Detectors of ``edge'' PPTES}

In the previous subsections we have have given two optimization
procedures. In both of them, starting from a general EW one can
obtain one which is optimal (or nd--optimal). It may well happen
that the EW found in this way is non--decomposable even though
the original one was decomposable. To check that one simply has
to use Corollary $3$; that is, check whether $W^T$ is an EW or
not. In case it is, then the OEW $W$ is non--decomposable.
However, nothing guarantees that the final EW is
non--decomposable if the original one is not. In this subsection
we describe a general method to construct nd--EW using the
optimization procedures introduced earlier. This method
generalizes the one presented in Ref.\ \cite{Te99}.

We are going to use the results presented in Ref.\
\cite{Kr99,Ho00}. There, we have already used and discussed the
``edge'' PPTES, without naming them, however. Let us now
introduce the following definition:

{\bf Definition:} [see Ref.\ \cite{Kr99}] A PPTES $\delta$ is an
``edge'' PPTES if for all product vector $|e,f\rangle$ and
$\epsilon>0$, $\delta-\epsilon |e,f\rangle\langle e,f|$ is not a
PPTES.

This definition implies that that the ``edge'' states lie on the
boundary between the PPTES and entangled states with
non--positive partial transpose. In this subsection we will show
how, out of an ``edge'' PPTES, we can construct a nd--OEW that
detects it. As we mentioned in the introduction, ``edge'' PPTES
are of special importance. In particular, they allow to provide
a canonical form to write PPTES in arbitrary Hilbert spaces. For
these reasons, some of the properties of the ``edge'' PPTES are
discussed in Appendix B.

In order to check whether a PPTES $\delta$ is an ``edge'' PPTES
we can use the range criterion \cite{Ho97} (see also
\cite{Kr99}). That is, $\delta$ is an ``edge'' PPTES iff for all
$|e,f\rangle\in R(\delta)$, $|e,f^\ast\rangle\not\in
R(\delta^{T_B})$.

Let $\delta$ be an ``edge'' PPTES, and let us denote by $P_1$
the projector onto $K(\delta)$ and by $Q_1$ the projector onto
$K(\delta^T)$. We define
\be
\label{Wd}
W_\delta=a(P_1 + Q_1^T),
\ee
where $a=1/\tr(P_1 + Q_1)$. Let us also define
\be
\label{epsilon1}
\epsilon_1\equiv \inf_{|e,f\rangle}\langle
e,f|W_{\delta}|e,f\rangle.
\ee
Then we have

{\bf Lemma 6:} Given an ``edge'' PPTES $\delta$, then
$W_1\propto W_\delta-\epsilon_{1}
\one$ is a nd--EW, where $\epsilon_1$ and $W_\delta$ are
defined in (\ref{epsilon1},\ref{Wd}), respectively.

{\em Proof:} We have that $\langle
e,f|W_\delta|e,f\rangle=a(\langle e,f|P_1|e,f\rangle + \langle
e,f^\ast|Q_{1}|e,f^\ast\rangle)\ge 0$. This quantity is zero iff
$\langle e,f|P_1|e,f\rangle=\langle
e,f^\ast|Q_{1}|e,f^\ast\rangle=0$. But this is not possible
since $\delta$ is an ``edge'' PPTES. Thus, $\langle
e,f|W_\delta|e,f\rangle>0$ for all $|e,f\rangle$. Defining
$\epsilon_{1}$ as in (\ref{epsilon1}), and taking into account
that $\langle e,f|W_\delta|e,f\rangle$ is a continuous function
of (the coefficients of) $|e,f\rangle$ and that the set in which
we are taken the infimum is compact, we obtain $\epsilon_1>0$.
Then we obviously have that $W_1$ fulfills properties (I) and
(III). On the other hand, $\bk{\delta}_{W_1}\propto
a(\bk{\delta}_{P_1}+ \bk{\delta^T}_{Q_1})-\epsilon_1 <0$, since
$P_1 \delta=Q_{1} \delta^T=0$. Thus, $W_1$ detects a PPTES, and
therefore, according to Theorem 3 is non--decomposable.$\Box$

Note that Lemma 6 provides an important generalization of the
method of Terhal \cite{Te99}, based on the use of unextendible
product bases \cite{Be98}. Our method works in Hilbert spaces of
arbitrary dimensions, and in particular when $\dim(H_A)=2$ (in
$2\times N$ dimensional systems) for which unextendible product
basis do not exist. By combining Lemma 6 and the optimization
procedure introduced earlier, we obtain a way of creating
nd--OEW. Once we have $W_1$ we find $p_{W_1}$ and $p_{W_1^T}$.
We denote by $P_2$ and $Q_2$ the projector operators orthogonal
to these two sets, respectively,
\be
\epsilon_2=\inf_{|e,f\rangle} \frac{\langle e,f|W_1|e,f\rangle}
{\langle e,f| P_2+Q_2^T|e,f\rangle},
\ee
and $W_2\propto W_1-\epsilon_2 (P_2+Q_2^T)$. According to Lemma
2b we have that $W_2$ is nd--finer than $W_1$. Now we can define
$p_{W_2},p_{W_2^T},P_3,Q_3$ and $W_3$ in the same way, and
continue in this vein until for some $k$, $\epsilon_k=0$. If
$W_k$ is not yet optimal, we still have to find other projectors
such that we can optimize as explained in the previous
subsections.

In Section VII we illustrate this method with a family of edge
PPTES from Ref. \cite{Ho97}. In fact, as we will mention in that
Section, we have checked that the optimization method typically
works as well by starting with three random vectors, and
following a similar procedure to the one indicated here. This
means that in our construction method we do not need in practice
to start from a ``edge'' PPTES.

\subsection{Sufficient condition for PPTES}

In this subsection we use the results derived in the previous
one to construct a sufficient criterion for non--separability of
PPTES. As shown in Ref. \cite{Kr99,Ho00}, given an operator
$\rho\ge 0$, with $\rho^T\ge 0$, we can always decompose it in
the form
\be
\label{decomposition}
\rho= \rho_s + \delta,
\ee
where $\rho_s$ is separable and $\delta$ is an ``edge'' PPTES.
More details concerning this decomposition, and in particular
its canonical optimal form are presented in Appendix B. In this
section we use this decomposition together with the following

{\bf Lemma 7:} Given a non--separable operator
$\rho=\rho_s+\delta$, where $\rho_s\ge 0$ is separable then for
all EW, $W$, such that $\bk{\rho}_W<0$ we have that
$\bk{\delta}_W<0$.

{\em Proof:} Obvious from the definition of EW. $\Box$

Lemma 7 tells us that if $\rho$ is non--separable, then there
must exist some EW that detect both $\delta$ and $\rho$.
Actually, it is clear that there must exist an OEW with that
property. In particular, if $\rho^T\ge 0$, it must be a nd--OEW.
In the previous subsection we have shown how to build them out
of ``edge'' PPTES. Thus, given $\rho$ we can always decompose it
in the form (\ref{decomposition}), construct an OEW that detects
$\delta$ and check whether it detects $\rho$. In that case, we
will have that $\rho$ is non--separable. Thus, this provides a
sufficient criterion for non--separability.

We stress the fact that for PPTES only a special class of
states, namely the class of ``edge'' PPTES, is responsible for
the entanglement properties. In fact, one should stress that
very many of the examples of PPTES discussed so far in the
literature belong to the class of ``edge'' PPTES: the $2 \otimes
4$ family from \cite{Ho97}, the $n
\otimes n$ states obtained via unextendible product basis construction
\cite{Be98}, the $3 \otimes 3$ states obtained via the
chess-board method \cite{more}(b), and projections of continuous
variable PPTES onto finite dimensional subspaces \cite{more}(c).

\section{Positive maps}

It is known that PM allow for necessary and sufficient conditions
for separability (or, equivalently, entanglement) of bipartite mixed
states \cite{Ho96}. PM's have been also applied in the context of 
distillation of entanglement \cite{xor} and information theoretic analysis
of separability \cite{Ce99}. 
In this Section we will use the isomorphism between operators
and linear maps to extend the properties derived for witnesses
to PM \cite{Ja72}. We will first review some of the definitions
and properties of linear maps.

Let us consider a linear map ${\cal E}: B(H_A)\to B(H_C)$. We
say that ${\cal E}$ is positive if for all $Y\in B(H_A)$
positive, ${\cal E}(Y)\ge 0$. One can extend a linear map as
follows. Given ${\cal E}: B(H_A)\to B(H_C)$, we define its
extension ${\cal E}\otimes 1_B: B(H_A)\otimes B(H_B)\to
B(H_C)\otimes B(H_B)$ according to ${\cal E}\otimes 1_C(\sum_i
Y_i\otimes Z_i)= \sum_i {\cal E}(Y_i)\otimes Z_i$, where $Y_i\in
B(H_A)$ and $Z_i\in B(H_B)$. A linear map is completely positive
if all extensions are positive. The classification and
characterization of positive (but not completely positive) maps
is an open question (see, e.g. Ref.\ \cite{Wo76,Ch82}).

An example of positive (but not completely positive) map is
transposition (in a given basis $O_A$); that is, the map ${\cal
E}_{T}$ such that ${\cal E}_T(Y)=Y^T$. The corresponding
extension is the partial transposition \cite{pt}. A map ${\cal
E}$ is called decomposable if it can be written as ${\cal
E}={\cal E}_{1} + {\cal E}_{2}\cdot {\cal E}_T$, where ${\cal
E}_{1,2}$ are completely positive.

One can relate linear maps with linear operators in the
following way. We will assume $d_A\equiv{\rm dim}(H_A)\le {\rm
dim}(H_C)$, but one can otherwise exchange $H_A$ by $H_C$ in
what follows. Given $X\in B(H_A\otimes H_C)$ and an orthonormal
basis $O_A=\{|k\rangle\}_{k=1}^{d_A}$ in $H_A$, we define the
linear map ${\cal E}_X: B(H_A)\to B(H_C)$ according to
\be
\label{EY}
{\cal E}(Y)=\tr_A(X^{T_A} Y),
\ee
for all $Y\in B(H_A)$, where $\tr_A$ denotes the trace in $H_A$
and the partial transpose is taken in the basis $O_A$.
Similarly, given a linear map we can always find an operator
$X$ such that (\ref{EY}) is fulfilled. For instance, if we
choose $T=(|\Psi\rangle\langle \Psi|)^{T_A}$, where
\be
|\Psi\rangle=\sum_{k=1}^{d_A}|k\rangle_A\otimes |k\rangle_C,
\ee
then the corresponding map ${\cal E}_T$ is precisely the
transposition in the basis $O_A$.

Given a linear map ${\cal E}_X$, one can easily show the
following relations: (a) ${\cal E}_X$ is completely positive iff
$X\ge 0$; (b) ${\cal E}_X$ is positive but not completely
positive iff $X$ is an EW [except for the normalization
condition (III)]; (c) ${\cal E}_X$ is decomposable iff $X$ is
decomposable. Thus, the problem of studying and classifying PM
is very much related to the one of EW. Furthermore, PM can be
also used to detect entanglement \cite{Ho96}. Let us consider
the extension $\bar {\cal E}_X\equiv {\cal E}_X\otimes 1:
B(H_A)\otimes B(H_B)\to B(H_C)\otimes B(H_B)$, where we take
$d_B\equiv {\rm dim}(H_B)= {\rm dim}(H_C)$. Then we have that
given $\rho\in B(H_A\otimes H_B)$,
\be
\bk{\rho}_X = \langle \Psi|\bar{\cal E}_X(\rho)|\Psi\rangle,
\ee
where
\be
|\Psi\rangle=\sum_{k=1}^{d_B}|k\rangle_C\otimes |k\rangle_B.
\ee
Thus, if an EW, $W$, detects $\rho$, then $\bar{\cal E}_W(\rho)$
is a non--positive operator. Consequently, $\rho\ge 0$ is
entangled iff there exists a PM such that acting on $\rho$ gives
a non--positive operator. In that case we say that the PM
``detects'' $\rho$. Actually, PM are ``more efficient'' in
detecting entanglement than EW. The reason is that it may happen
that $\bar{\cal E}_X(\rho)$ is non--positive but still
$\bk{\rho}_X\ge 0$.

It is convenient to define finer and optimal PM as for EW. That
is, given two PM, ${\cal E}_{1,2}$, we say that ${\cal E}_2$ is
finer than ${\cal E}_1$ if it detects more. We say that a PM,
${\cal E}$, is optimal if there exists no one that is finer. In
the same way we can define nd--finer and nd--optimal.

The results presented in the previous sections can be directly
translated to PM given the following fact.

{\bf Lemma 8:} If $W_2$ is finer (nd--finer) than $W_1$ then
${\cal E}_{W_2}$ is finer (nd--finer) than ${\cal E}_{W_1}$.

{\em Proof:} Using Lemma 2 we can write
$W_1=(1-\epsilon)W_2+\epsilon P$. According to (\ref{EY}) we
have that ${\cal E}_{W_1}=(1-\epsilon){\cal E}_{W_2}+\epsilon
{\cal E}_P$. Since ${\cal E}_P(\rho)\ge 0$ for all $\rho\ge 0$,
we have that ${\cal E}_{W_2}$ is finer than ${\cal E}_{W_1}$.
Using Lemma 2b we can also prove that it is nd--finer. $\Box$

>From this lemma it follows that optimizing EW implies optimizing
PM. In fact, the constructions that we have given in the Section
VC can be viewed as ways of constructing non--decomposable PM.
In fact, since the method works for $\dim(H_A)=2$, the resulting
PM ${\cal E}: B(H_A)\to B(H_C)$ has a minimal ``qubit'' domain,
or -- equivalently -- minimal hermitian conjugate codomain. Up
to our knowledge, our method is the first one that permits to
construct non--decomposable PM with these characteristics.

\section{Illustration}

In this section we explicitly give construct a nd--OEW out of
edge PPTES. We use, as a starting point, the family of PPTES
introduced in \cite{Ho97}).

\subsection{Family of ``edge'' PPTES}

We consider $H_A=\C^2$ and $H_B=\C^4$, and denote by
$\{|k\rangle\}_{k=0}^{d_\alpha}$ ($\alpha=A,B$) an orthonormal basis in
these spaces, respectively. Most of the time we will write the operators
in those bases; that is, as matrices. For operators acting in
$H_A\otimes H_B$ we will always use the following order
$\{|0,0\rangle,|0,1\rangle,\ldots, |1,0\rangle,|1,1\rangle,\ldots\}$. On
the other hand, all partial transposes will be taken with respect to
$H_B$.

We consider the following family of positive operators
\cite{Ho97}
\bea
\label{rhob}
\rho_b=\frac{1}{7b+1}\left( \begin{array}{cccccccc}
b&0&0&0&0&b&0&0\\
0&b&0&0&0&0&b&0\\
0&0&b&0&0&0&0&b\\
 0&0&0&b&0&0&0&0\\
0&0&0&0&\frac{1+b}{2}&0&0&\frac{\sqrt{1-b^2}}{2}\\
b&0&0&0&0&b&0&0\\
0&b&0&0&0&0&b&0\\
 0&0&b&0&\frac{\sqrt{1-b^2}}{2}&0&0&\frac{1+b}{2}\\
\end{array}\right),
\eea
where $b\in[0,1]$. For $b=0,1$ those states are separable,
whereas for $0<b<1$, $\rho_b$ is an ``edge'' PPTES. This can be
shown by checking directly that they violate the range criterion
of Ref. \cite{Ho97}, i.e. the definition given in Section IVC.

If we take the partial transpose in the basis $\{|k\rangle\}$,
the density operators $\rho_b$ have the property that
$\rho_b^{T}=U_B \rho_b U_B^\dagger$ with $U_B=(\sigma_x)_{03}
\oplus (\sigma_x)_{12}$. Here, the subscript $ij$ denotes the
subspace, ${\cal H}_{Bij}\subset{\cal H}_{B}$ spanned by
$\{|i\rangle, |j\rangle\}$ and $\sigma_x$ is one of the
Pauli-operators. Note that $U_B$ is a real unitary operator
acting only on $H_B$. This immediately implies that
\be
\label{property}
\tilde{\rho}_b^{T}=\tilde\rho_b,
\ee
where $\tilde{\rho}_b= V_B \rho_b V_B^\dagger$ and
$V_B=1\sqrt{2}[(\one+i\sigma_x)_{03}\oplus
(\one+i\sigma_x)_{12}]$. We will use the property
(\ref{property}) to simplify the problem of constructing the
nd--OEW. Thus, we will concentrate from now on the operators
$\tilde\rho_b$ \cite{changebasis}. Obviously, $\tilde\rho_b$ is
an ``edge'' PPTES for $1>b>0$.

The projector onto the kernel of $\tilde\rho_b$, $P_1$, is
invariant under the transformation $T_{AB}=T_A\otimes T_B$, where
\bea
T_A &=& \left( \begin{array}{cc}
 1&0\\ 0&e^{i 2\pi/3}\\
\end{array}\right), \\
T_B&=&\left( \begin{array}{cccc}
 1&0&0&0\\
0&\cos{(2\pi/3)}&-\sin{(2\pi/3)}&0\\
 0&\sin{(2\pi/3)}&\cos{(2\pi/3)}&0\\
0&0&0&1\\ \end{array}\right).\nonumber
\eea
Note that $T_B$ is a real matrix. Later
on we will need its eigenstates with real coefficients; they are
$|0\rangle\pm|3\rangle$. Note also that $T_{AB}^3=\one$.

\subsection{Construction of nd--EW's}

We use now the methods developed in Section V to obtain a
nd--OEW starting from $\tilde \rho_b$. That is, we define
$W_b=P_1+P_1^{T}$, where $P_1$ is the projector onto
$K(\tilde\rho_b)=K(\tilde\rho_b^T)$. Our procedure consists of
first subtracting the identity to obtain
$W_1=W_b-\epsilon_1\one$. Then, we subtract $P_2+Q_2^T$,
$P_3+Q_3^T$, etc. In the n--th step we will have
\be
W_{n}=W_{n-1}-\epsilon_{n}(P_{n}+Q_{n}^T),
\ee
where $P_{n}$ ($Q_{n}$) is the projector orthogonal to the space
spanned by $P_{W_{n-1}}$ ($P_{W_{n-1}^T}$). We will use the symmetries of
$\tilde\rho_b$ to better understand the structure of $W_{n}$.

\begin{description}

\item[(a)] $W_{n}=W^{T}_n$. We can prove this by induction. First, it is
clear that $W_1=W_1^T$. Let us now assume that $W_{n-1}=W_{n-1}^T$. Then
we show that $W_{n}=W^{T}_n$. For that, we just have to show that the
subspace spanned by $P_{W_{n-1}}$ is the same that the one spanned by
$P_{W_{n-1}^T}$, so that $Q_{n}=P_n$. But this is clear since $W_{n-1}
=W_{n-1}^T$. $\Box$

\item[(b)] $T_{AB}W_n T_{AB}^\dagger = W_n$. We prove this by induction.
First, for $W_1=P_1+P_1^{T}-\epsilon_1\one$ we have that $T_{AB} W_1
T^{\dagger}_{AB} =T_{AB} P_1 T^{\dagger}_{AB}+T_{AB} P^{T}_1
T^{\dagger}_{AB}-\epsilon_1\one=W_1$, since $T_{AB} P^{T}_1
T^{\dagger}_{AB}= (T_{AB} P_1 T^{\dagger}_{AB})^T$ (given the fact that
$T_B$ is real) and $P_1$ is invariant under $T_{AB}$. Then, let us
assume that $T_{AB}W_{n-1} T_{AB}^\dagger = W_{n-1}$. In order to
show that $T_{AB}W_n T_{AB}^\dagger = W_n$ we just have to show that
$P_n$ is invariant under $T_{AB}$, or, equivalently, that the subspace
spanned by $P_{W_{n-1}}$ is invariant under $T_{AB}$. But this
follows immediately from the fact that $T_{AB}W_{n-1} T_{AB}^\dagger = W_{n-1}$.
$\Box$

\end{description}

Starting the property (a) it follows that the vectors
$|e,f\rangle\in P_{W_n}$ will have $|f\rangle$ real (unless we
have degeneracies). This can be seen by noticing that those
vectors minimize $\langle e,f |W_n|e,f\rangle$; defining $W_e
\equiv\langle e|W_n|e\rangle$, we have that
$W_e^T=W_e=W_e^\dagger$ is symmetric, and therefore the
eigenstate corresponding to its minimum eigenvalue can be chosen
to be real. On the other hand, starting from the property (b) it
follows that if $|e,f\rangle\in P_{W_n}$ then
$T_{AB}^\dagger|e,f\rangle, T_{AB}^{\dagger 2}|e,f\rangle\in
P_{W_n}$. According to that, we will typically have two kinds of
product vectors in $P_{W_n}$:

\begin{description}

\item[(1)] $|e,f\rangle$ is an eigenstate of $T^{\dagger}_{AB}$ with
$|f\rangle$ real: There are only $4$
possible product vectors which fulfill these conditions:
$\{|0\rangle,|1\rangle\} \otimes
\{|0\rangle+|3\rangle,|0\rangle-
|3\rangle \}$.

\item[(2)] $|e,f\rangle$ is not an eigenstate of $T^{\dagger}_{AB}$: Then,
we will also have: $T^{\dagger}_{AB}|e,f\rangle$ and
$(T^{\dagger}_{AB})^2|e,f\rangle \in P_W$.

\end{description}

We have carried out this procedure for $\tilde\rho_b$ and found
nd--OEW for each $b$. We find that for the optimal EW we have
two vectors of the kind (1) and six of the kind (2). In total we
find eight product vectors in $P_W$, which span the whole
Hilbert space and therefore the corresponding EW are optimal
(see Corollary 2). This means that any operator of the form
(\ref{rhow}) with $|e_k,f_k\rangle\in P_W$ the product vectors
we have found, and $p_k>0$ will be a full range separable
density operator that lies on the boundary between separable and
PPTES. Up to our knowledge, this constitutes the first example
of those operators \cite{comm0}. We have also created the PM corresponding to
the nd--OEW, which are the first examples of
non--decomposable PM with minimal ``qubit'' domain, or --
equivalently -- minimal hermitian conjugate codomain.

In Fig.\ 1 we show for which $b^\prime$ $\tilde\rho_{b^\prime}$
is still detected by the nd--OEW created out of
$\tilde\rho_{b}$. We find that for a given $b$, the optimal
witness that we create detects all $\tilde\rho_{\tilde b}$ for
$\tilde b\le b'$. Thus, in the figure we plot $b'$ as a function
of $b$. As explained above, the corresponding positive map
detects more than the witness itself. In the figure one can also
see how much is detected by the positive map.

\begin{figure}[ht]
\begin{picture}(230,200)
\put(5,5){\epsfxsize=230pt\epsffile[23 146 546 590 ]{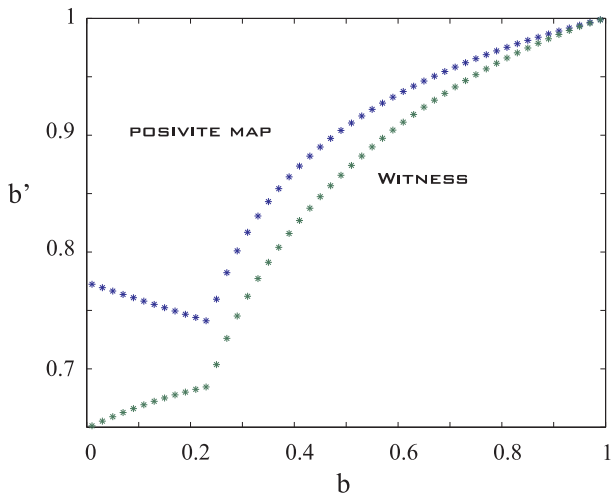}}
\end{picture}
\caption[]{Values of $b'$ for which if $\tilde b\le b'$, $\tilde\rho_{\tilde b}$
is detected by the witness and the positive map created starting from $\tilde\rho_b$.}
\label{Fig1}
\end{figure}

Obviously, the witnesses that we create do not only detect the
density operators $\tilde\rho_b$. For instance one can check how
much one can add the identity to certain $\tilde \rho_b$ but
still keeping the state entangled. That is for which $\lambda$,
$\tilde\rho_b+\lambda\one$ is still detected by the witness.
This is shown in the following figure.

\begin{figure}[ht]
\begin{picture}(230,200)
\put(5,5){\epsfxsize=230pt\epsffile[51 191 567 615]{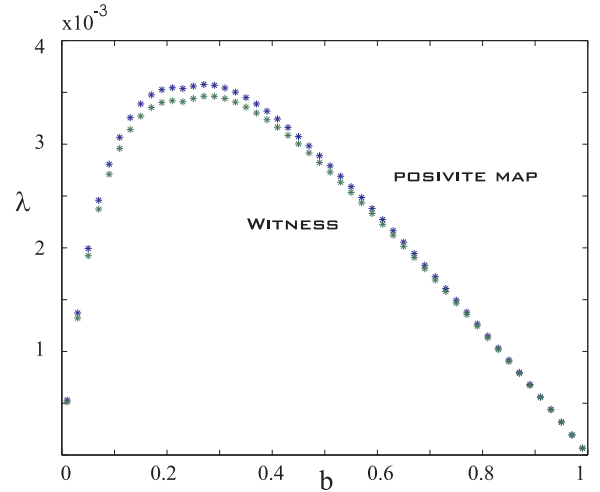}}
\end{picture}
\caption[]{Maximum $\lambda$ such that $\tilde\rho_b+\lambda \one$ is
still detected by the witness and the positive map created starting
from $\tilde\rho_b$.}
\label{Fig2}
\end{figure}

Finally, let us note that we have observed using numerical calculations
that if one starts with a random projector, $P$ of rank $3$, and
optimizes the decomposable operator $W\equiv P+P^{T_B}$ in the same way
as the one described here, then one will end up with a nd--OEW
$\tilde{W}$, where $p_{\tilde{W}}$ is complete. This means that in
order to create nd--OEW one does not need to know in practice an edge
PPTES. In another words, optimization itself is a way to reach
nondecomposableness.

\subsection{Analytical procedure}

In this subsection we will present an analytical way to create nd--EW's.
Furthermore we will given an example of such a witness, which detects
$\rho_b$ for all $b\in(0,1)$. From Fig. $1$ we see that the witness which
detects most is the one we created out of $\tilde\rho_b$, where $b$ is very
close to $1$. We will work with the original $\rho_b$ (\ref{rhob}).

We consider two
hermitian operators $A$ and $B$, with $A$ positive on product vectors, i.e.,
$\langle e,f|A|e,f\rangle\geq 0$, whereas $B$ does not have to. As
before we denote by $P_A$ ($P_B$) the (not necessarily complete) set of
product vectors on which $A$ ($B$) vanishes. We require that for all
$|e,f\rangle \in P_A$, $\langle e,f|B|e,f\rangle\geq 0$. Then we define
$W(x)\equiv \frac{1}{x}(A+xB)$ for any real $x$. So we have the
following

{\bf Lemma 9:} If $\mbox{lim}_{x\rightarrow 0}\bk{\rho}_{W(x)}
<0$ then $\rho$ is entangled.

{\em Proof:} We prove that $\mbox{lim}_{x\rightarrow 0}\langle
e,f|W(x) |e,f\rangle \geq 0$. This implies that if $\rho$ is
separable, then $\mbox{lim}_{x\rightarrow 0}\bk{\rho}_{W(x}\geq 0$.
 Let us therefore distinguish two cases: (i)
if $|e,f\rangle \in P_A$ then we have that
$\mbox{lim}_{x\rightarrow 0}\langle e,f|W(x) |e,f\rangle
=\langle e,f|B|e,f\rangle$, which is, per assumption, positive.
(ii) $|e,f\rangle \not \in P_A$ then we have
$\mbox{lim}_{x\rightarrow 0}\langle e,f|W(x) |e,f\rangle
=\mbox{lim}_{x\rightarrow 0}\frac{a}{x}+b$, where $a=\langle
e,f|A|e,f\rangle > 0$ and $b=\langle e,f|B|e,f\rangle$. Thus
this limit tends to infinity, which proves the statement.$\Box$

Note that $W(x)$ is not an EW since it is not necessarily  positive on
product vectors. However, one can make it positive by adding the identity
operator to convert it into an EW.

{\bf Corollary 6:} Given any $x_0 >0$, then $W(x_0)\equiv
\frac{1}{x_0}(A+x_0 B)+\lambda_{x_{0}} \one$ , with
$\lambda_{x_{0}}=-\mbox{ min}_{|e,f\rangle} \langle
e,f|\frac{1}{x_0}(A+x_0 B) |e,f\rangle$ is an EW.

Let us now illustrate how we can use Lemma 9 to detect all the
states $\rho_b$. We define
\bea
A=\left( \begin{array}{cccccccc}
0&0&0&0&0&0&0&0\\
0&1&0&0&0&0&-2&0\\
0&0&1&0&0&0&0&0\\
 0&0&0&0&0&0&0&0\\
 0&0&0&0&0&0&0&0\\
0&0&0&0&0&1&0&0\\
0&-2&0&0&0&0&1&0\\
0&0&0&0&0&0&0&0\\
\end{array}\right),
\eea
\bea
B=\left( \begin{array}{cccccccc}
1&0&0&1&0&-2&0&0\\
0&1&0&0&0&0&0&0\\
0&0&1&0&0&0&0&-2\\
1&0&0&1&0&0&0&0\\
0&0&0&0&1&0&0&-1\\
-2&0&0&0&0&1&0&0\\
0&0&0&0&0&0&1&0\\
0&0&-2&0&-1&0&0&1\\
\end{array}\right).
\eea

One can easily show that $A=|\psi\rangle \langle \psi|+(|\phi\rangle
\langle \phi|)^{T_B}$, where $|\psi\rangle =|01\rangle-|12\rangle$ and
$|\phi\rangle=|02\rangle-|11\rangle$. Thus this operator is positive on
product vectors, since it is decomposable. Let us now use unnormalized
states in order to present the set of product vectors on which $A$
vanishes, i.e. $P_A$. $P_A=P_{A_{1}}\cup P_{A_{2}}$, where
$P_{A_{1}}=\{(|0\rangle +\alpha |1\rangle)\otimes
(x|0\rangle+y|3\rangle) \forall \alpha ,x,y\}$ and
$P_{A_{2}}=\{(|0\rangle +e^{i\Phi} |1\rangle)\otimes
[x|0\rangle+y|3\rangle+z(|1\rangle+e^{-i\Phi} |2\rangle] \forall
\Phi,x,y,z\}$. The operator $B$ has to be positive on those product
vectors, i.e., $\forall |e,f\rangle \in P_A$, $\langle
e,f|B|e,f\rangle\geq 0$. In order to show that this is indeed
like that let us distinguish the two cases: $|e,f\rangle \in
P_{A_{1}}$ and $|e,f\rangle \in P_{A_{2}}$. In the first case we
have that
\bea
\langle e|B|e\rangle =\left( \begin{array}{cccc}
1+|\alpha|^2&-2\alpha&0&1-|\alpha|^2\\
-2\alpha^\ast&1+|\alpha|^2&0&0\\
0&0&1+|\alpha|^2&-2\alpha\\
1-|\alpha|^2&0&-2\alpha^\ast&1+|\alpha|^2
\end{array}\right)
\eea
 and so $\langle e,f|B|e,f\rangle =|x+y|^2+|\alpha|^2|x-y|^2 \geq 0$. If
$|e,f\rangle \in P_{A_{2}}$ then
\bea
\langle e|B|e\rangle =\left( \begin{array}{cccc}
2&-2e^{i\Phi}&0&0\\
-2e^{-i\Phi}&2&0&0\\
0&0&2&-2e^{i\Phi}\\
0&0&-2e^{-i\Phi}&2
\end{array}\right)
\eea
which is a positive operator and so $\langle e,f|B|e,f\rangle
\geq 0$. So those two operators $A$ and $B$ fulfill all the required
properties. Furthermore one can show that $\bk{\rho_b}_A=0$ and
$\bk{\rho_b}_B<0$ for all $0<b<1$. Thus we have that
$\mbox{lim}_{x\rightarrow 0}\bk{W(x)
\rho_b}
<0$ for all $0<b<1$, where we defined
$W(x)=\frac{1}{x}(A+xB)$.

As mentioned above we can use now $W(x)$ in order to create
other PPTES just by adding product vectors on which $W(x)$
vanishes. To find the product vectors we can add, all we need to
do is to determine the intersection between $P_A$ and $P_B$.
Since $P_B=\{(|0\rangle +e^{i\phi} |1\rangle)\otimes
[a(|0\rangle+e^{-i\phi}|1\rangle+b(|2\rangle+e^{-i\phi}
|3\rangle)]
\forall \phi,a,b\}$ we have that $S\equiv P_A\cap P_B=P_{A_2}\cap
P_B=\{(|0\rangle +e^{i\phi} |1\rangle)\otimes
(|0\rangle+e^{-i\phi}|1\rangle+e^{-i2\phi}|2\rangle+e^{-
i3\phi}|3\rangle) \forall \phi\}$. Note that $S$ spans a $5$ dimensional
subspace and that the orthogonal subspace is spanned by the vectors
$\{-|02\rangle+|13\rangle,
-|01\rangle+|12\rangle,-|00\rangle+|11\rangle\}$.

\section{Conclusions}

Entanglement witnesses allow us to study the separability
properties of density operators. We have defined OEW, which are
those that detect entanglement in an optimal way. We have given
necessary and sufficient conditions for an EW to be optimal, and
we have shown a way to construct them. We have also concentrated
on nd--EW, which are those that detect PPTES. We have extended
the definitions of optimality and the optimization procedure to
those EW. It turns out that one can optimize nd--EW by
subtracting decomposable operators. We have also given an
explicit method to construct nd--EW starting from ``edge''
PPTES. We have also mentioned that this method works by starting
out from random operators. We have extended our techniques to
PM, and therefore given a method to systematically construct
non--decomposable positive maps. We have illustrated our methods
with a family of ``edge'' PPTES acting on $\C^2\otimes\C^4$. The
corresponding PM constitute the first examples of PM with minimal
``qubit'' domain, or -- equivalently -- minimal hermitian
conjugate codomain. We have also constructed the first examples
of separable states of full range that lie on the boundary
between separable and PPTES. These states can be used for 
experimental realization of PPTES \cite{comm0}.

In this paper we have also introduced the ``edge'' PPTES, which
violate the range criterion of separability. As shown in
Appendix B, the ``edge'' PPTES allow us to construct a canonical
form of PPTES in Hilbert spaces of arbitrary dimensions. They
also allow us to give a novel sufficient condition for
non--separability which applies to operators with positive
partial transpose. It is based on the fact that among all PM (or
EW) only the subset $\{\Lambda_{edge}\}$ of those PM that detect
edge PPTES are needed to study the separability of PPTES. This
opens many interesting questions. Is it possible that in the set
$\{\Lambda_{edge}\}$ there is some map that is globally finer
than the transposition? In another words, is there a map
detecting the entanglement of {\em all} the states with
non--positive partial transpose? What is the minimal subset of
$\{\Lambda_{edge}\}$ providing such condition? Is it finite?

Finally, let us consider the implications of the our results for
the very interesting problem of locality of PPTES. There is a
conjecture \cite{LHVPT} that those states can be local in the
sense that they admit a local hidden variable (LHV) model for
any set of possible local measurements. The problem is not
trivial given the fact that it may be important to take into
account the role of sequential measurements and the possible
existence of many copies. Quite recently it has been shown that
PPTES satisfy Bell-type of inequalities introduced by Mermin
\cite{BellPT}. It is not difficult to convince oneself that the
set of states admitting LHV model for {\it any} fixed type of
measurements is a convex set. Furthermore, extending the
reasoning from \cite{We89} it is easy to see that the set of
separable states admits LHV models for any possible set of
measurements. Hence, taking into account the results of this
paper it follows that in order to prove, or to disprove locality
of PPTES it is enough to study only ``edge'' PPTES.

Note that the ``edge'' states have typically very small rank
(the minimal rank is four in $3\otimes3$ systems, see Ref.\
\cite{Ho00}). There have been no examples of LHV models for
states of low rank, so far. Thus, perhaps completely new
techniques will be needed to study this problem. In this case
the most symmetric PPTES provided recently \cite{more}(c) seem
to be the best suitable for the first test.

\section{Acknowledgments}

This was has been supported in part by the Deutsche
Forschungsgemeinschaft (SFB 407 and Schwerpunkt
"Quanteninformationsverarbeitung"), the DAAD, the Austrian
Science Foundation (SFB ``control and measurement of coherent
quantum systems''), the ESF PESC Programm on Quantum
Information, TMR network ERB--FMRX--CT96--0087, the IST
Programme EQUIP, and the Institute for Quantum Information GmbH.

\appendix

\section{Optimality of EW}

In this Appendix we study necessary and sufficient conditions
for an EW to be optimal. According to Theorem 1 of Section III
we have that an EW, $W$, is optimal iff no positive operator can
be subtracted from $W$ while keeping property (I). This
condition can be reexpressed in terms of the infimum of some
scalar products in Lemma 4. This infimum is, in general,
difficult to calculate (at least analytically). In this Section
we will give a different method to determine whether an EW is
optimal or not. This method will turn out to be very simple for
the case in which $\dim(H_A)=2$. The idea is to find the
conditions such that a given operator $P\ge 0$ can/cannot be
subtracted from an EW. This will give us automatically a
criterion to determine when $W$ is optimal.

In all this appendix we will use that given an EW, $W$, and an
operator $P\ge 0$ we say that $P$ cannot be subtracted from $W$
if for all $\lambda>0$, $W-\lambda P$ does not fulfill (I). In
other words, there exist $|e(\lambda)\rangle\in H_A$ and
$|f(\lambda)\rangle\in H_B$ such that
\be
\label{bor}
\langle e(\lambda),f(\lambda)|(W-\lambda P)|e(\lambda),f(\lambda)\rangle
< 0.
\ee
Note that $\langle
e(\lambda),f(\lambda)|P|e(\lambda),f(\lambda)\rangle$ must be
strictly positive, so that (\ref{bor}) can be expressed as
\be
\label{bor2}
\lim_{\lambda\to 0} \frac{\langle
e(\lambda),f(\lambda)|W|e(\lambda),f(\lambda)\rangle} {\langle
e(\lambda),f(\lambda)|P|e(\lambda),f(\lambda)\rangle}=0.
\ee

In the first subsection we will introduce some definitions and notation.
In the second one we give a method to determine the set of product
vectors $P_W$, on which $W$ vanishes. In the third subsection we find a
necessary and sufficient condition under which an operator cannot be
subtracted from an EW. We will see that there must exist a vector
$|e_0,f_0\rangle\in P_W$, some other vectors $|e_1\rangle$ and
$|f_1\rangle$, and certain phases $\phi_{e,f}$ and $\theta$ such that
some quantity is zero. In the next subsection we will see that the
problem can be reduced to finding only the vectors $|e_{0,1}\rangle$ and
$|f_{0,1}\rangle$. Finally, we will show that if $\dim(H_A)=2$ we just
have to find $|e_0\rangle$ and $|f_0\rangle$, which is very simple.

\subsection{Definitions and notation}

In order to prove the results of this appendix in a compact and readable
form we have made an extensive numbers of definitions.

We will always denote by $|e_0,f_0\rangle$ a product vector in $P_W$,
and by $|e_1\rangle\in H_A$ and $|f_1\rangle\in H_B$ two vectors
orthogonal to $|e_0\rangle$ and $|f_0\rangle$, respectively. We will use
the following notation:
\be
W_{i,j}^{k,l} = \langle e_i,f_j|W|e_k,f_l\rangle, \quad
(i,j,k,l=0,1).
\ee
and we will write
\begin{mathletters}
\bea
W_{1,0}^{0,1} &=& |W_{1,0}^{0,1}|e^{i\phi_0}\\ W_{0,0}^{1,1} &=&
|W_{0,0}^{1,1}|e^{i\phi_1}.
\eea
\end{mathletters}
We will also define the following operators:
\begin{mathletters}
\bea
w_{i,j}^e &\equiv& \langle e_i|W|e_j\rangle,\\
 w_{i,j}^f
&\equiv&
\langle f_i|W|f_j\rangle.
\eea
\end{mathletters}

The following vectors will be used in the context of Eq.\
(\ref{bor2}):
\begin{mathletters}
\label{ht}
\bea
|e(\epsilon)\rangle &=&
\frac{1}{\sqrt{1+|\cos(\theta)\epsilon|^2}}
 (|e_0\rangle+\epsilon \cos(\theta) e^{i\phi_e}|e_1\rangle),\\
|f(\epsilon)\rangle &=&
\frac{1}{\sqrt{1+|\sin(\theta)\epsilon|^2}}
 (|f_0\rangle+\epsilon \sin(\theta)e^{i\phi_f}|f_1\rangle),
\eea
\end{mathletters}
where $\epsilon$ is a real number, and $\phi_{e,f}\in [0,\pi)$
and $\theta\in [0,\pi/2]$ are certain constants. Given a product
vector $|e(\epsilon),f(\epsilon)\rangle$ and an operator, $W$,
we will expand $\langle
e(\epsilon),f(\epsilon)|W|e(\epsilon),f(\epsilon)\rangle$ by
collecting terms with the same powers in $\epsilon$; that is,
except for a normalization constant,
\be
\label{Aexpan}
\langle e(\epsilon),f(\epsilon)|W|e(\epsilon),f(\epsilon)\rangle
\propto \sum_{i=1}^4 \epsilon^i A_i(W),
\ee
where
\begin{mathletters}
\label{AAA}
\bea
 A_0(W) &=& W_{0,0}^{0,0},\\
 A_1(W) &=& 2 {\rm Re}\left[ \cos(\theta) e^{i\phi_e} W_{0,0}^{1,0}
       + \sin(\theta) e^{i\phi_f} W_{0,0}^{0,1}\right],\\
\label{A2}
 A_2(W) &=&  \cos^2(\theta) W_{1,0}^{1,0} + \sin^2(\theta) W_{0,1}^{0,1}\\
        & &+ 2 \sin(\theta)\cos(\theta) \nonumber\\
        & & \times {\rm Re}\left[e^{-i(\phi_e-\phi_f)} W_{1,0}^{0,1} +
       e^{i(\phi_e+\phi_f)}  W_{0,0}^{1,1}\right] ,\nonumber \\
 A_3(W) &=& 2 \sin(\theta)\cos(\theta) \\
        & & \times {\rm Re}\left[\cos(\theta) e^{i\phi_f} W_{1,0}^{1,1} +
       \sin(\theta) e^{i\phi_e} W_{0,1}^{1,1} \right],\nonumber \\
 A_4(W) &=&  \sin^2(\theta) \cos^2(\theta) W_{1,1}^{1,1}.
\eea
\end{mathletters}
On the other hand, we will define
\be
\label{PSI}
|\Psi_{0,1}\rangle\equiv\sin(\theta)e^{i\phi_f}|e_0,f_1\rangle+\cos(\theta)
e^{i\phi_e}|e_1,f_0\rangle.
\ee

Finally, the following quantity will play an important role in
determining whether there exist vectors and parameters for which
(\ref{bor2}):
\be
\label{XW}
X(W)\equiv W_{1,0}^{1,0}W_{0,1}^{0,1} -
(|W_{1,0}^{0,1}|+|W_{0,0}^{1,1}|)^2.
\ee

\subsection{Determining $P_W$}

As stated in Lemma 3, not every positive operator $P$ can be
subtracted from an EW, $W$; it must vanish on $P_W$. Thus, in
order to choose $P$ one has to know the set $P_W$. In this
subsection we give a method to determine it.

We start by characterizing the vectors in $P_W$:

{\bf Lemma A1:} Given an operator $W$ satisfying (I), then
$|e_0,f_0\rangle\in P_W$ iff
\begin{mathletters}
\label{eqq}
\bea
\label{eqa}
\langle e_0|W|e_0\rangle |f_0\rangle &=&0,\\
\label{eqb}
\langle f_0|W|f_0\rangle |e_0\rangle &=&0,
\eea
\end{mathletters}

{\em Proof:} (If) We just apply $\langle f_0|$ to Eq.\
(\ref{eqa}). (Only if) Since $W$ fulfills (I) then
$W_{e_0}\equiv \langle e_0|W|e_0\rangle$ must be positive. Thus,
$\langle f_0|W_{e_0}|f_0\rangle=0$ implies Eq.\ (\ref{eqa}). In
the same way we obtain Eq.\ (\ref{eqb}). $\Box$

In practice, for a given $W$ the set $P_W$ can be found as follows. Due
to the fact that $W$ is an EW we have that for any $|e\rangle \in H_A$,
$W_e\equiv\langle e|W|e\rangle$ must be a positive operator (i.e.
$\langle f|W_e|f\rangle\ge 0$ for all $|f\rangle\in H_B$). Thus, the
determinant ${\rm det}(W_e)\ge 0$. According to Lemma A1, this
determinant is zero iff there exists some $|f_0\rangle\in H_B$ such that
$\langle f_0|W_{e_0}|f_0\rangle=0$, i.e., if $|e_0,f_0\rangle\in
P_W$. That is, the determinant as a function of $|e\rangle$ has a
minimum (which is zero) at $|e_0\rangle$. We can use this fact
to find $|e_0\rangle$. Then, we can easily obtain $|f_0\rangle$ via Eq.\ (\ref{eqa}).
We can expand an unnormalized state
$|e\rangle$ in an orthonormal basis $\{|k\rangle\}$ as
\be
|e\rangle = \sum_{k=1}^{\dim(H_A)} c_k |k\rangle,
\ee
and impose that the corresponding determinant is zero. This gives us a
polynomial equation for the coefficients $c_k$, i.e.
\be
P(c_k,c_k^\ast)=0.
\ee
We also impose that, given the fact that the determinant is a
minimum,
\be
\frac{\partial}{\partial c_k} P(c_k,c_k^\ast)=
\frac{\partial}{\partial c_k^\ast} P(c_k,c_k^\ast)=0,
\ee
which also give a set of polynomial equations. These equations
can be solved using the method mentioned in Ref.\ \cite{Kr99}.

\subsection{Necessary and sufficient conditions for subtracting
an operator}

In this subsection we give a necessary and sufficient condition
for an operator $P$ to be subtractable from an EW. We start out
by giving some properties of the coefficients $A(W)$ defined
above (\ref{AAA}).

{\bf Lemma A2:} Given $W$ satisfying (I) and $|e_0,f_0\rangle\in
P_W$, then for all $|e_1\rangle\in H_A$ and $|f_1\rangle\in H_B$
we have

\begin{description}
\item [(i)] $A_0(W)=A_1(W)=0$.

\item [(ii)] $A_2(W)\ge 0$.

\item [(iii)] If $A_2(W)=0$ then $A_3(W)=0$.
\end{description}

{\em Proof:} (i) It is a direct consequence from Lemma A1. In
order to prove (ii--iii) we use the fact that $W$ satisfies (I).
We define $|e(\epsilon)\rangle$ and $|f(\epsilon)\rangle$ as in
(\ref{ht}). We impose that $\langle
e(\epsilon),f(\epsilon)|W|e(\epsilon),f(\epsilon)\rangle\ge 0$.
Using the expansion (\ref{Aexpan}) and taking into account (i),
we have $A(\epsilon)\equiv A_2(W) + \epsilon A_3(W) + \epsilon^2
A_4(W) \ge 0$ for all $\epsilon$. This automatically implies
(ii), since otherwise for sufficiently small $\epsilon$ we would
have $A(\epsilon)<0$. It also implies (iii), since if $A_3(W)<0$
($A_3(W)>0$) then for sufficiently small $\epsilon>0$
($\epsilon<0$) we would have $A(\epsilon)<0$. $\Box$

Now, we are at the position of giving a necessary and sufficient
condition under which an operator cannot be subtracted from an
EW:

{\bf Lemma A3:} Given $P$ fulfilling $P P_W=0$, it cannot be subtracted
from $W$ iff there exists $|e_0,f_0\rangle\in P_W$, $|e_1\rangle\perp
|e_0\rangle$, $|f_1\rangle\perp |f_0\rangle$, $\phi_{e,f}$, and $\theta$
such that $A_2(W)=0$ but $A_2(P)\ne 0$.

{\em Proof:} (If) We define $|e(\lambda)\rangle$ and
$|f(\lambda)\rangle$ as in (\ref{ht}). Using Lemma A2(i) we have
$A_0(W)=A_0(P)=A_1(W)=A_1(P)=0$. Using Lemma A2(iii) we have
that $A_3(W)=0$. Thus, we can write the limit (\ref{bor2}) as
\be
\label{lim}
\lim_{\lambda\to 0} \frac{
\lambda^2 A_4(W)}
{A_2(P)+\lambda A_3(P)+\lambda^2 A_4(P)}
\ee
which obviously tends to zero given that $A_2(P)\ne 0$. (Only
if) There exist two normalized vectors $|\tilde
e(\lambda)\rangle$ and $|\tilde f(\lambda)\rangle$ (continuous
functions of $\lambda$) fulfilling (\ref{bor2}). Taking the
limit $\lambda\to 0$ in this expression we have that $\langle
\tilde e(0),\tilde f(0)|W|\tilde e(0),\tilde f(0)\rangle=0$,
and therefore $|e_0,f_0\rangle
\equiv |\tilde e(0),\tilde f(0)\rangle\in P_W$.
This means that we can always choose $|\tilde
e(\lambda)\rangle=|e[\epsilon(\lambda)]\rangle$ and $|\tilde
f(\lambda)\rangle=|f[\epsilon(\lambda)]\rangle$ given in
(\ref{ht}), where $|e_1\rangle\perp |e_0\rangle$ and
$|f_1\rangle\perp |f_0\rangle$ are two normalized vectors,
$\lim_{\lambda\to 0} \epsilon(\lambda)=0$, and $\langle
e(\epsilon),f(\epsilon)|P|e(\epsilon),f(\epsilon)\rangle\ne 0$.
We use (\ref{ht}) to expand the numerator and denominator of
(\ref{bor2}) as in (\ref{Aexpan}). According to Lemma A2(i) we
have that $A_0(W)=A_0(P)=A_1(W)=A_1(P)=0$. Thus, we must have
\be
\label{lim2}
\lim_{\epsilon\to 0} \frac{
A_2(W)+\epsilon A_3(W)+\epsilon^2 A_4(W)} { A_2(P)+\epsilon
A_3(P)+\epsilon^2 A_4(P)} =0.
\ee
This implies $A_2(W)=0$ and $A_2(P)\ne 0$. Note that if both
$A_2(W)=A_2(P)=0$ then, according to Lemma A2(iii) we have that
$A_3(W)=A_3(P)=0$, so that (\ref{lim}) would require
$A_4(W)/A_4(P)=0$. But this cannot be since $A_4(W)=0$ would
imply that $|e(\epsilon),f(\epsilon)\rangle \in P_W$, and
therefore $\langle
e(\epsilon),f(\epsilon|P|e(\epsilon),f(\epsilon)\rangle=0$.
$\Box$

Finally, we show in the next lemma that condition $A_2(P)=0$ is
equivalent to having certain vector in the kernel of $P$. We
will use the vector $|\Psi_{0,1}\rangle$ defined in (\ref{PSI}).

{\bf Lemma A4:} Given a positive operator $P$ and a set of
vectors $|e_0,f_0\rangle\in K(P)$, $|e_1\rangle\perp
|e_0\rangle$ $|f_1\rangle\perp |f_0\rangle$, and parameters
$\phi_{e,f}$, and $\theta$ then $A_2(P)=0$ iff
$|\Psi_{0,1}\rangle \in K(P)$.

{\em Proof:} Since $P\ge 0$ and $|e_0,f_0\rangle\in K(P)$ we
have $P_{1,1}^{0,0}=0$. Then, we can write $A_2(P)=\langle
\Psi_{0,1}|P|\Psi_{0,1}\rangle$, with $|\Psi\rangle$ is defined in
(\ref{PSI}), from which it is obvious that $A_2(P)=0$ iff
$|\Psi_{0,1}\rangle\in K(P)$. $\Box$

\subsection{Necessary and sufficient conditions for $A_2(W)=0$}

The previous lemmas tell us that we cannot subtract a given
operator $P$ provided we can find some vectors and parameters
such that $A_2(W)=0$. The task of finding these vectors is
difficult, in general. Here we will give a way to check whether
these vectors exist. As before, we will denote by
$|e_0,f_0\rangle$ a vector in $P_W$, and by $|e_1\rangle$ and
$|f_1\rangle$ two vectors orthogonal to the first two. The
quantity $X(W)$ defined in (\ref{XW}) will play an important
role in determining whether there exist vectors and parameters
for which $A_2(W)=0$. In this subsection, we will always have to
choose the phases $\phi_{e,f}$ that minimize $A_2(W)$. That is
\be
\label{phis}
e^{-i(\phi_e-\phi_f-\phi_0)}=-1, \quad
e^{i(\phi_e+\phi_f+\phi_1)}=-1.
\ee
We will denote $\tilde A_2(W)$ the value of $A_2(W)$ for this
particular choice of phases. We have
\bea
\label{ineq}
\tilde A_2(W) &=&
\cos^2(\theta) W_{1,0}^{1,0} +
\sin^2(\theta)  W_{0,1}^{0,1} \\
 && -2 \sin(\theta)\cos(\theta)
 \sqrt{W_{1,0}^{1,0}W_{0,1}^{0,1}- X(W)},\nonumber
\eea
where we have used (\ref{XW}).

Let us start showing that $X(W)$ is positive. We will use this
property later on to reexpress the condition $A_2(W)=0$ in terms
of one that is simpler to check.

{\bf Lemma A5:} $X(W)\ge 0$.

{\em Proof:} This follows from the fact that $A_2(W)\ge 0$ for
all values of $\phi_{e,f}$. In particular, $\tilde A_2(W)\ge 0$,
which according to (\ref{ineq}) implies $X(W)\ge 0$. $\Box$

The next lemma shows that we just have to check whether $X(W)=0$
if we want to see if there exist parameters $\phi_{e,f}$ and
$\theta$ such that $A_2(W)=0$. This first condition is therefore
much more useful than the last one.

{\bf Lemma A6:} $X(W)=0$ iff there exist $\phi_{e,f}^0$ and
$\theta^0$ such that $A_2(W)=0$.

{\em Proof:} (If) Given the phase $\theta=\theta^0$ we have that
$0=A_2(W)\ge \tilde A_2(W)$. Thus, $\tilde A_2(W)=0$. According to
(\ref{ineq}) we can have two cases: (a) $\theta_0\ne 0,\pi/2$.
In that case it is obvious that $X(W)=0$. (b)
$\theta_0=0,\pi/2$. In the first (second) case we must have
$W_{1,0}^{1,0}=0$ ($W_{0,1}^{0,1}=0$). But this implies that
$W_{0,0}^{1,1}= W_{0,1}^{1,0}=0$ since otherwise we could always
find some other value of $\theta$ such that $\tilde A_2(W)<0$.
Then, $X(W)=0$. (Only if) We choose $\phi_{e,f}$ as in
(\ref{phis}). For this value, according to (\ref{ineq}) we have
\be
\label{caca}
\tilde A_2(W)= \left[\cos(\theta)\sqrt{W_{1,0}^{1,0}}
- \sin(\theta)\sqrt{W_{0,1}^{0,1}}\right]^2,
\ee
which can always be zero for some particular value of $\theta$.
$\Box$

Note that according to the proof of Lemma A6, if $W_{1,0}^{1,0}=0$ then
$A_2(W)=0$ only for $\theta=0$. But in that case one can easily check
that the vector $|e(\lambda),f(\lambda)\rangle\in P_W$ [see (\ref{ht})]
which cannot be. Similarly, we conclude that $W_{1,0}^{1,0}\ne 0$ if we
want $A_2(W)=0$. Thus, from now one we will assume that both
$W_{1,0}^{1,0}$ and $W_{1,0}^{1,0}$ are not zero.

\subsection{Optimality test}

Thus, we can now state the steps to check whether an EW, $W$,
can be optimized or not. (1) For each $|e_0,f_0\rangle\in P_W$
we must check whether there exist $|e_1\rangle \perp
|e_0\rangle$ and $|f_1\rangle\perp |f_0\rangle$ such that
$X(W)=0$. Let us denote by $|e_{0,1}^{(i)}\rangle$ and
$|f_{0,1}^{(i)}\rangle$ the set of vectors fulfilling that. (2)
For each of these vectors, we have to find the corresponding
values of $\phi_{e,f}^{(i)}$ by using (\ref{phis}) and of
$\theta^{(i)}$ by imposing that $\tilde A_2(W)=0$ in
(\ref{caca}). (3) Construct $|\Psi^{(i)}\rangle$ according to
(\ref{PSI}). (4) See whether the space spanned by $P_W$ and
$\{|\Psi^{(i)}\rangle\}$ is equal to $H_A\otimes H_B$. If it is,
then $W$ is optimal. If it is not, we can always find some
$|\psi\rangle$ orthogonal to that subspace that can be
subtracted from $W$.

\subsection{Necessary and sufficient conditions for $X(W)=0$}

The hard part of the procedure outlined before to see whether and EW is
optimal is the step (1), namely to find $|e_1\rangle$ and $|f_1\rangle$
such that $X(W)=0$. We start out by giving a necessary and sufficient
condition for $X(W)=0$.

{\bf Lemma A7:} Given $|e_0,f_0\rangle\in P_W$, and
$|e_1\rangle\perp |e_0\rangle$ and $|f_1\rangle\perp
|f_0\rangle$, then $X(W)=0$ iff
\begin{mathletters}
\label{cogno2}
\bea
\label{congo3}
w_{0,0}^e|f_1\rangle &=&
- \sqrt{\frac{W_{0,1}^{0,1}}{W_{1,0}^{1,0}}} e^{-i\phi_f}(e^{-i\phi_e}w_{1,0}^e
+ e^{i\phi_e} w_{0,1}^e)|f_0\rangle,\\
\label{congo4}
w_{0,0}^f|e_1\rangle &=&
- \sqrt{\frac{W_{0,1}^{0,1}}{W_{1,0}^{1,0}}} e^{-i\phi_e}(e^{-i\phi_f} w_{1,0}^f
+ e^{i\phi_f} w_{0,1}^f)|e_0\rangle,
\eea
\end{mathletters}
where $\phi_{e,f}$ are given in (\ref{phis}).

{\em Proof:} (If) We multiply by $\langle f_1|$ Eq.\
(\ref{congo3}) and take the square of the absolute value of the
result. We obtain
\bea
W_{1,0}^{1,0}W_{0,1}^{0,1} &=& |e^{-i(\phi_e+\phi_f)} W_{1,1}^{0,0} +
e^{i(\phi_e-\phi_f)} W_{0,1}^{1,0}|^2\nonumber\\
&\le& (|W^{1,1}_{0,0}| + |W_{1,0}^{0,1}|)^2.
\eea
Using Lemma A5 we conclude that $X(W)=0$. (Only if) Since $X(W)=0$ and
according to Lemma A5 $X(W)\ge 0$, then $X(W)$ must be a minimum
with respect to $|e_1\rangle$ and $|f_1\rangle$. Taking the
derivatives of $X(W)$ with respect to these two vectors and
imposing that they vanish, one obtains (\ref{cogno2}). $\Box$

Equations (\ref{cogno2}) are particularly useful if the dimension of one
of the Hilbert spaces is 2. Without loss of generality, let us assume
that $\dim(H_A)=2$. In that case we can choose $|e_1\rangle$ as the one
that is orthogonal to $|e_0\rangle$ (with an arbitrary choice of the
global phase). The determination of $\phi_e$ can be done as follows.
Using (\ref{cogno2}) we write
\be
\label{f1}
\sqrt{\frac{W_{1,0}^{1,0}}{W_{0,1}^{0,1}}}e^{i\phi_f} |f_1\rangle =
- \frac{1}{w_{0,0}^e} (e^{-i\phi_e} w_{1,0}^e + e^{i\phi_e}
w_{0,1}^e)|f_0\rangle
\ee
where $1/w_{0,0}^e$ denotes the pseudo--inverse
\cite{notepseudoinv}. We can use this expression to impose
\be
\label{WWW}
W_{1,0}^{0,1} e^{-i(\phi_e-\phi_f)},W_{0,0}^{1,1} e^{i(\phi_e+\phi_f)}<0,
\ee
i.e. they are negative real numbers. We obtain that
\be
\label{condi}
e^{-i2\phi_e} \langle f_0| w_{1,0}^e \frac{1}{w_{0,0}^e} w_{1,0}^e|f_0\rangle <0,
\ee
so that we determine $\phi_e$. With these results, we can prove
the following necessary and sufficient condition for $X(W)=0$
when $\dim(H_A)=2$.

{\bf Lemma A8:} If $\dim(H_A)=2$, given $|e_0,f_0\rangle\in
P_W$, then there exists $|e_1,f_1\rangle$ such that $X(W)=0$ iff
\bea
\label{cond2}
& &\langle f_0| \left[w_{1,1}^e - w_{0,1}^e
\frac{1}{w_{0,0}^e}w_{1,0}^e - w_{1,0}^e
\frac{1}{w_{0,0}^e}w_{0,1}^e\right] |f_0\rangle =\nonumber\\
&& 2 \Big|
\langle f_0|w_{0,1}^e \frac{1}{w_{0,0}^e}w_{0,1}^e|f_0\rangle
\Big|.
\eea

{\em Proof:} (If) We define
\be
|f_1\rangle =
- \frac{1}{w_{0,0}^e} (e^{-i\phi_e} w_{1,0}^e + e^{i\phi_e}
w_{0,1}^e)|f_0\rangle
\ee
where $\phi_e$ is determined by the condition (\ref{condi}). Using
this expression to calculate $X(W)$ one finds that indeed $X(W)=0$.
(Only if) Using Lemma A7 we can write $|f_1\rangle$ as in (\ref{f1})
so that the phases $\phi_{e,f}$ ensure that (\ref{WWW}) is fulfilled.
Substituting $|f_1\rangle$ in the equation
$X(W)=0$ one finds (\ref{cond2}). $\Box$

In summary, for a given $|e_0,f_0\rangle\in P_W$, in order to find
whether there exist $|e_1,f_1\rangle$ such that $X(W)=0$ we just have to
check the condition (\ref{cond2}). If it is fulfilled, we can easily
find $|f_1\rangle$ and the phases $\phi_{e,f}$ using (\ref{WWW}) and
(\ref{f1}).

\section{Canonical form of PPTES}

The concept of ``edge'' PPTES seems to play a very special role
in the characterization of PPTES. In particular, in view of the
criterion given in Section VD, which is based on the fact that
any density operator $\rho$ can be decomposed into a separable
part and an ``edge'' PPTES (\ref{decomposition}). Among all the
possible decompositions there might be one for which the trace
of the separable part is maximal. When it exists, such a
decomposition was termed positive partial transpose best
separable approximation (PPT BSA) to $\rho$ \cite{Ho00}. It
extended the idea of BSA introduced in Refs.\ \cite{Sa98,Le98}
to the case of PPTES, which were based on the method of
diminishing the range of $\rho$ by subtracting product vectors
from its range, while keeping the remainder and, at the same
time, its partial transpose, positive
\cite{Sa98,Le98,Kr99,Ho00}. In this Appendix we formalize the
results regarding the existence and properties of the PPT BSA.
In particular, the proofs presented in the quoted papers were
restricted to the case in which there exist a finite, or at
most, countable number of projectors on product vectors that can
be subtracted from $\rho$. We will extend them below to
continuous families of product vectors. The Appendix is written
in a self-contained way, and can be read independently of the
body of the paper.

We denote by $\Gamma_{\rho}$ the set of projectors on product
vectors $\{|e_{\alpha}, f_{\alpha}\rangle \langle e_{\alpha},
f_{\alpha}|\}$ such that $|e_{\alpha}, f_{\alpha}\rangle \in
R(\rho)$ and $|e_{\alpha}, f_{\alpha}^* \rangle \in
R(\rho^{T_{B}})$. In Ref.\ \cite{Ho00} we showed that if
$\Gamma_\rho$ is finite then there exist an optimal
decomposition (PPT BSA) $\rho=(1-p) \rho_{sep} + p \delta $
where $\delta$ is an ``edge'' PPTES, and $p$ is minimal. Note
that PPT BSA involves the state $\delta$ which violates the
range criterion in a rather special way, i.e. with the
additional requirement that $\Gamma_\rho$ is a finite set. It
can happen that there is an uncountable family of product
vectors depending on continuous parameter that can be used for
subtracting projectors. In the following we will show that in
such case the above result is valid.

In order to consider the case of continuous families of product vectors
we first prove the following:

{\bf Lemma B1:} Let $\rho$ will be a PPTES defined on a Hilbert
space ${\cal H}$ , ${\rm dim} {\cal H}<\infty$. Then the set of
product vectors $\Gamma_{\rho}$ is compact.

{\em Proof :} Obviously $\Gamma_{\rho}$ is a bounded set in
finite-dimensional space, so it is enough to show that it is
closed. Consider any sequence $|g_{n},h_{n} \rangle \rightarrow
| \phi\rangle$, $|g_{n},h_{n} \rangle \in R(\rho)$,
$|g_{n},h_{n}^{*} \rangle \in R(\rho^{T_{B}})$. The limit vector
must: (i) respect the condition of orthogonality to $K(\rho)$
[i. e. they must belong to $R(\rho)$], (ii) belong to the sphere
(i. e. set of all vectors $|\phi\rangle$ with $||\phi||=1$),
(iii) finally, it must be a product state, because if it was
entangled then its distance from the compact set of product pure
states\cite{Ho97} defined as $\min_{|e,f\rangle} |||\phi\rangle
- |e,f\rangle||$ would be nonzero, which is obviously
impossible. We conclude thus $|\phi\rangle =|g,h\rangle \in
R(\rho)$ for some $|g\rangle,|h \rangle$, which implies (up to
irrelevant phase factors) that $|g_{n}\rangle \rightarrow
|g\rangle$ and $|h_{n}\rangle
\rightarrow |h \rangle$. We have (again up to irrelevant external phase
factor) $| g_{n}, h_{n}^*\rangle \rightarrow |g, h^*\rangle$.
The latter must belong to $R(\rho^{T_{B}})$ as any element of
the corresponding sequence is orthogonal to $K(\rho^{T_{B}})$.
$\Box$

Let us now prove the following general lemma, which is a generalization
of one theorem from Ref. \cite{Le98}:

{\bf Lemma B2:} Let the PPTES $\rho$ be defined on a finite
dimensional Hilbert space. Consider the set $\Sigma_{\rho}$
consisting of the trivial zero operator plus all unnormalized
states $\tilde{\rho}$ ($\tr\tilde{\rho}\leq 1$) such that
$\tilde{\delta}\equiv\rho -
\tilde{\rho}$ is positive and has positive partial transpose. Then, one can find
$\hat{\rho} \in \Sigma_\rho$ such that with $\tr(\hat{\rho})\leq 1$ is
optimal in the sense that:

\begin{description}

\item [(i)] The trace of
$\hat{\delta}\equiv \rho-\hat{\rho}$ is minimal with respect to
all separable $\tilde{\rho}$'s leading to positive partial
transpose $\tilde{\delta}$'s.

\item [(ii)] The state $\delta=\hat\delta/\tr(\hat\delta)$
is an ``edge'' PPTES.

\end{description}

{\em Proof :} To prove the existence of $\hat{\rho} \in
\Sigma_{\rho}$ we just have to show that $\Sigma_{\rho}$ is compact.
This can be done by showing that $\Sigma_{\rho}$ is a closed subset of
another compact set, namely $C={\rm conv} \{ \Gamma_{\rho} \cup$ {\bf
0} $ \} $. The latter set $C$ is compact as it is a convex hull of the
compact set $\{ \Gamma_{\rho} \cup$ {\bf 0} $ \} $ in a finite
dimensional space.

Note first that $\Sigma_{\rho} \subset C$. Indeed, by virtue of $\tilde{\delta}
\geq 0$ any nonzero $\tilde{\rho}$ cannot have any vector in its range not
belonging to $R(\rho)$. Analogously
$R(\tilde{\rho}^{T_{B}})\subset R(\rho^{T_{B}})$. Hence,
according to the properties of the ranges of density operators
in general \cite{Ho97}, $\tilde{\rho}$ must be a convex
combination of vectors from $\Gamma_{\rho}$, and as such it
belongs to $C$. Let us show that $\Sigma_{\rho}$ is closed. This
follows immediately form the fact that $\Sigma_\rho$ is a
cross-section (performed over any projections $P$, $Q$) of the
sets: $\Sigma^{1}_{\rho,P}\equiv \{ \tilde{\rho}:
f_{P,\rho}(\tilde{\rho})\equiv \tr(P\rho - P\tilde{\rho})\geq 0
\}$ and $\Sigma^{2}_{\rho,Q}=\{ \tilde{\rho}:
g_{Q,\rho}(\tilde{\rho})\equiv \tr(Q^{T_B}\rho - Q^{T_B}
\tilde{\rho})
\geq 0
\}$. Since the functions $f_{P,\rho}, g_{Q,\rho}$ are continuous, 
all the sets participating in the cross section are closed.
Now, the cross-section of closed sets is again a closed one.

Consider now the statement (ii). Since $\delta, \delta^{T_{B}}\geq 0$,
we always have $\delta=\beta P_{R(\delta)}+A$ and $\delta^{T_{B}}=\beta'
P_{R(\delta^{T_{B}})}+A'$ with $\beta, \beta' >0$, some positive
operators $A, A'$ (here, $P_{X}$ denotes a projector onto the subspace
$X \subset {\cal H}$). Then if, contrary to (ii), there were any
$|e,f\rangle \in R(\delta)$ such that $|e,f^{*}\rangle \in
R(\delta^{T_{B}})$, then the new operator
$\hat{\rho}^*=\hat{\rho}+\gamma|e, f\rangle\langle e, f|$,
$\gamma={\rm min}[\beta, \beta']$ would fulfill that
 $\hat{\delta}^*=\rho-\hat{\rho}^*$ is a PPTES, and would
contradict optimality with respect to (i). $\Box$

Let us remark that if we give up the condition regarding positivity of 
$\tilde{\delta}^{T_{B}}$, then we obtain a modified statement (ii) where
the state $\delta$ has no product vectors in its range. This is nothing
but the best separable approximation (BSA) of Ref. \cite{Le98}, extended
here rigorously to the states $\rho$ having uncountable set of product
vectors in $R(\rho)$.

From the Lemma B2 we obtain  the following characterization
of PPTES, which can be regarded to be among the main
results of this appendix, since it provides
{\it a canonical form of PPTES}:

{\bf Proposition :}
If the state $\rho$ is PPTES, then it is a convex combination
\be
\label{decomp}
\rho=(1-p) \rho_{sep} + p \delta
\ee
of some normalized separable $\rho_{sep}$ and a normalized ``edge'' PPTES $\delta$.
In the above decomposition the weight $p$ is minimal [i. e.
there does not exist a decomposition of type (\ref{decomp}) with
a smaller $p$].

The above proposition means, in particular, that
{\it the edge PPTES are responsible for
PPT type entanglement}.



\begin{references}
\bibitem{pt}
Given an operator $X$ and an orthonormal basis
$\{|k\rangle\}_{k=1}^N\in H_B$, one defines the partial
transposed of $X$ with respect to $B$ in that basis as follows:
\be
X^{T_B} = \sum_{k,k'=1}^N |k'\rangle_B\langle
k|X|k'\rangle_B\langle k|.
\ee
One can analogously define the partial transposed of $X$ with
respect to $A$ in a given basis, $X^{T_A}$. Partial
transposition fulfills the following useful property
\be
\label{PropPT}
\tr(X^{T_B} Y)=\tr(XY^{T_B}).
\ee
We say that positive operator has a positive partial
transposition if $\rho^{T_B}\ge 0$. Note that this property is
basis independent, and that $\rho^{T_B}\ge 0$ iff $\rho^{T_A}\ge
0$. The relation between separable and positive partial
transpose operators was established in Refs.\ \cite{Pe96,Ho96}.
All separable operators have positive partial transposition. The
converse is, however, not true in general. That is, there are
positive partial transpose operators which are nonseparable.
\bibitem{note0}
This normalization condition expresses the fact that if $W$
fulfills (I--II), and $\lambda$ is a positive number, then
$\lambda W$ also fulfills (I--II), so that we can always
normalize $W$ to have (III). Given an operator satisfying (I)
and (II) we can always define $W'=W/\tr(W)$, so that it fulfills
both properties as well and detects the same operators as $W$.
This follows from the fact that $\tr(W)$ must be strictly
positive since if we take the trace in an orthonormal basis of
product vectors, using (I) we have that $\tr(W)\ge 0$. If
$\tr(W)=0$ then we would have that $\langle e,f|W|e,f\rangle=0$
for all product vectors; since we can always construct a
complete set of operators $\{|e_i,f_i\rangle\langle e_i,f_i|\}$
out of product vectors in the space of the operators acting on
$H_A\otimes H_B$, this immediately implies that $W=0$, which
contradicts (II). This set of operators can be constructed as
follows. Let us consider a Hilbert space $H_A$ of dimension
$d_a$ and an orthonormal basis $\{|k\rangle\}_{k=1}^{d_a}$. We
define the states $|\Psi^{r}_{k,k'}\rangle=(|k\rangle +
|k'\rangle)/\sqrt{2}$ and $|\Psi^{i}_{k,k'}\rangle=(|k\rangle +
i |k'\rangle)/\sqrt{2}$ for $k< k'$ and the projectors on these
states $A_k=|k\rangle\langle k|$,
$A_{k,k'}^{r}=|\Psi^r_{k,k'}\rangle\langle\Psi^r_{k,k'}|$, and
$A_{k,k'}^{i}=|\Psi^i_{k,k'}\rangle\langle\Psi^i_{k,k'}|$. The
set $S_A=\{A_k, A_{k,k'}^{r}, A_{k,k'}^{i}, k=1,\ldots,d_a,
k'=k+1,\ldots,d_a\}$ is complete in the space of operators
acting on $H_A$. This can be easily shown by noting that
$|k\rangle\langle k'|=
A_{k,k'}^{r}+iA_{k,k'}^{r}-(1+i)(A_k+A_{k'})/2$ for $k\ne k'$.
We can do the same construction for $H_B$ and obtain a set
$S_B$. The set composed of tensor products of all elements of
$S_A$ with $S_B$ is complete in the space of operators acting on
$H_A\otimes H_B$.
\bibitem{pt2}
Whenever it does not matter whether we take the partial
transpose of an operator $X$ with respect to the first or second
system we will simply write $X^T$.
\bibitem{notepseudoinv}
Given a selfadjoint operator $X$, its pseudoinverse $X^{-1}$ is
defined as the one that vanishes in $K(X)$ and so that
$XX^{-1}=X^{-1}X$ is the projector in $R(X)$. If $X$ has no zero
eigenvalue, the pseudoinverse coincides with the inverse
operator.
\bibitem{changebasis}
Note that if we define a basis in $H_B$ as
$\{V_B^\dagger|k\rangle\}$ we have that $\rho_b^{T}=\rho_b$,
where the partial transposition is taken in that basis.
\bibitem{comm0} These states can be prepared locally and, 
can be used to construct the PPTES by mixing them ``weakly'' 
with entangled states. This provides an interesting possibility
of experimental realization of PPTES. This suggestion has been
formulated by A. Weinfurter.
\end{references}
\end{document}